\documentclass[a4paper,11pt]{article}
\pdfoutput=1 
\usepackage{paper_jheppub}
\usepackage{subfigure}
\usepackage{tabularx}
\usepackage[T1]{fontenc}

\def\GeV{\ifmmode {\mathrm{\ Ge\kern -0.1em V}}\else \textrm{Ge\kern -0.1em V}\fi}%

\newcommand{\DeltarelpT}{{\Delta^{\text{rel}} _{\pT}}}

\title{\boldmath Search for Non-Standard Sources of Parity Violation in Jets at $\sqrt{s}=8$ TeV with CMS Open Data}

\author[a]{Christopher G.~Lester}
\author[b,c]{Matthias Schott}

\affiliation[a]{Cavendish Laboratory, University of Cambridge, UK}
\affiliation[b]{Massachusetts Institute of Technology, Cambridge, USA}
\affiliation[c]{Johannes Gutenberg-University, Mainz, Germany}

\newcommand{\pT}{\ensuremath{p_{\mathrm{T}}}}

\newcommand{\ET}{\ensuremath{E_{\mathrm{T}}}}
\newcommand{\MET}{\mbox{\ensuremath{\not \!\! \ET}}}

\def\ifb{\mbox{fb$^{-1}$}}
\def\ipb{\mbox{pb$^{-1}$}}
\def\GeV{\ifmmode {\mathrm{\ Ge\kern -0.1em V}}\else \textrm{Ge\kern -0.1em V}\fi}%

\newcommand{\namea}{j_1}
\newcommand{\nameb}{j_2}
\newcommand{\namec}{j_3}
\newcommand{\named}{j_4}
\newcommand{\moma}{{\vec p^{\namea}}}
\newcommand{\momb}{{\vec p^{\nameb}}}
\newcommand{\momc}{{\vec p^{\namec}}}

\emailAdd{lester@hep.phy.cam.ac.uk, matthias.schott@cern.ch}

\abstract{
The Standard Model violates parity, but only by mechanisms which are invisible to Large Hadron Collider (LHC) experiments (on account of the lack of initial state polarisation or spin-sensitivity in the detectors).  Nonetheless, new physical processes could potentially violate parity in ways which {\em are} detectable by those same experiments.  If those sources of new physics occur only at LHC energies, they are untested by direct searches. We probe the feasibility of such measurements using approximately 0.2~\ifb\ of data which was recorded in 2012 by the CMS collaboration and made public within the CMS Open Data initiative. In particular, we test an inclusive three-jet event selection which is primarily sensitive to non-standard parity violating effects in quark-gluon interactions.  Within our measurements, no significant deviation from the Standard Model is seen and no obvious experimental limitations have been found. We discuss other ways that searches for non-standard parity violation could be performed, noting that these would be sensitive to very different sorts of models to those which our method would constrain. We hope that our initial studies provide a valuable starting point for rigorous future analyses using the full LHC datasets at 13 TeV with a careful and less conservative estimate of experimental uncertainties.}

\begin{document}

\maketitle

\addtocontents{toc}{\protect\setcounter{tocdepth}{1}} 

\section{\label{Sec:Intro}Introduction}

It is rightly beyond doubt that the laws of physics violate parity.  The elegant experiments of the 1950s \cite{Wu:1957my,Garwin:1957hc,PhysRev.109.1015} unambiguously showed that the weak interaction of the Standard Model can tell the difference between our universe and its mirror image.
However, no attempts have yet been made to identify whether there are non-standard parity violation mechanisms which operate only at high energies probed  by the Large Hadron Collider (LHC).

The absence of tests of parity is largely a pragmatic response to the obstacles presented by the LHC:  its beams are not polarised,  its detectors are not sensitive to polarizations, and it is mathematically impossible to construct a parity violating spin-averaged matrix element within any $CP$-conserving Locally Lorentz-Invariant quantum field theory (LLIQFT), effective or otherwise.\footnote{To reveal the existence of parity-violation a model must possess at least one matrix element having both a parity-even and a parity-odd part.
After trace identities have removed spinor sums, the only parity-odd expressions which can remain in a Lorentz-invariant $|M|^2$ are contractions of the totally antisymmetric alternating tensor with groups of four linearly independent four-momenta: $\epsilon_{\mu\nu\sigma\tau}a^\mu b^\nu c^\sigma d^\tau$.  \label{fn:epsilon} While such terms are parity-odd, they are also time-odd. 
Assuming $CPT$-symmetry, such a matrix element therefore also violates $CP$.  $CP$-conserving local Lorentz-invariant quantum field theories therefore cannot generate  parity violating differential cross sections. 
}
Given that the only route to probing parity-violation in the Standard Model at the LHC would be from within the $C$-even part of its (very small) $CP$-violating sector,\footnote{One could, in principle, demonstrate parity-violation unambiguously by using a `genuine $CP$-odd' observable (such as one of those described in \cite{Han:2009ra})
 on a $C$-even final state.} it is not surprising that no such analyses have yet been performed.\footnote{This limitation does not prevented the LHC from making measurements of parity-violating parameters within models {\it in which a particular mechanism of parity-violation is present by assumption}.
For example, the differences between the axial and vector couplings of the $Z$-boson in the Standard Model violate parity and were  measured in \cite{Aad:2015uau,Khachatryan:2016yte}.  However, neither of these papers incontrovertibly demonstrates that nature violates parity

The reason is simple:  the angles from which forward-backward asymmetries are calculated are even under parity,  unlike primary observables from the experiments of the 1950s.   
 The very same forward backward asymmetries could therefore also be explained, at least in principle, by some alternative parity {\it conserving} theory.
}

It would be wrong, however, to conclude that genuine tests of parity are therefore of limited value.
On the contrary, we will see that genuine tests of parity-invariance are very straightforward to make and can (in principle) provide arbitrarily large signatures for models not fitting into the class of LLIQFTs.

While it is true that there is an overwhelming theoretical preference for LLIQFTs (the Standard Model itself is one, as are most popular extensions including those featuring supersymmetry, leptoquarks, technicolor, axions, additional gauge interactions, {\it etc}.) there is no law of nature which {\it  demands} that  new physics be describable only by such theories.

Moreover, given the lack of evidence for new physics found at LHC thus far, the need for the community to search in all possible hiding places is surely greater than ever.  In particular, it is hard to imagine any reason why every possible attempt should not be made to test and re-test the fundamental symmetries of nature every time a door opens onto a new energy range.

We use therefore CMS Open Data to make the first LHC search for unequivocal evidence of parity violation. Clearly, the possible physics reach of our study is limited - once by the data statistics used, secondly by the assumed systematic uncertainties; its primary goal is to evaluate potential experimental challenges, while the actual constrains on parity violation are of secondary importance. The Standard Model itself should provide no appreciable signal on account of its approximate $CP$-symmetry. While sensitivity might exist to LLIQFT extensions of the Standard Model having larger sources of $CP$-violation, these are probably constrained by data from rare decays. The primary sensitivity, thus, is to extensions which are outside the LLIQFT class altogether.

The paper is structured as follows:
Section~\ref{Sec:Intro} (this introduction) has explained why genuine tests of parity violation in LHC data are important, and discusses the  limitations which the LHC imposes on which sources of parity violation are observable;
Section~\ref{Sec:ParityViolation} discusses the properties of the specific parity-odd event-variable, $\alpha$, used in this paper;
Section~\ref{Sec:CMS} summarises the CMS detector, its relevant reconstruction objects, and the data-sets used in this analysis;
Section~\ref{Sec:Calibration} discusses the calibration of the relevant physics objects used (in particular jets) used in this analysis;
Section~\ref{sec:Analysis} details the actual analysis strategy, signal selection and background estimation;
Section \ref{sec:ResultsAndDatasetAsymmetry} both defines the test statistic $A_\alpha$ which is used to provide the a measure of non-standard parity violation in the dataset as a whole, and then shows the asymmetries seen in CMS Open Data; 
Section~\ref{sec:Correction} unfolds the preceding results for detector effects.
The paper concludes in Section \ref{Sec:Conclusion}.

\section{\label{Sec:ParityViolation}Jet-based Signatures for Non-Standard Parity Violation}

One of the evident disadvantages of searching for the breaking of a particular symmetry without a particular model in mind, is that there the number of potential signatures is almost unbounded.  If testing for genuine parity violation,  any parity-odd variable which is invariant under the action of relevant non-parity symmetries\footnote{For example: an ideal parity-testing variable should be invariant under rotation of an LHC event by any angle about the beam pipe axis, or by 180 degrees about any axis at right angles to the beam pipe, unless sensitivity to Lorentz violating effects is desired.}  has the capacity 
(assuming a parity-even event selection)
to be sensitive to \textit{some} parity-violating model, albeit not necessarily one with a strong theoretical motivation.

Lacking clear guidance from any well motivated non LLIQFT models, we  make no claims of generality or optimality concerning our choice to concentrate on a signal region containing three jets. We argue only that the investigation is interesting for the reasons already given in the introduction.

\begin{figure}[t]
\centering
\begin{minipage}{7.3cm}
  \centering
\includegraphics[width=1.0\linewidth]{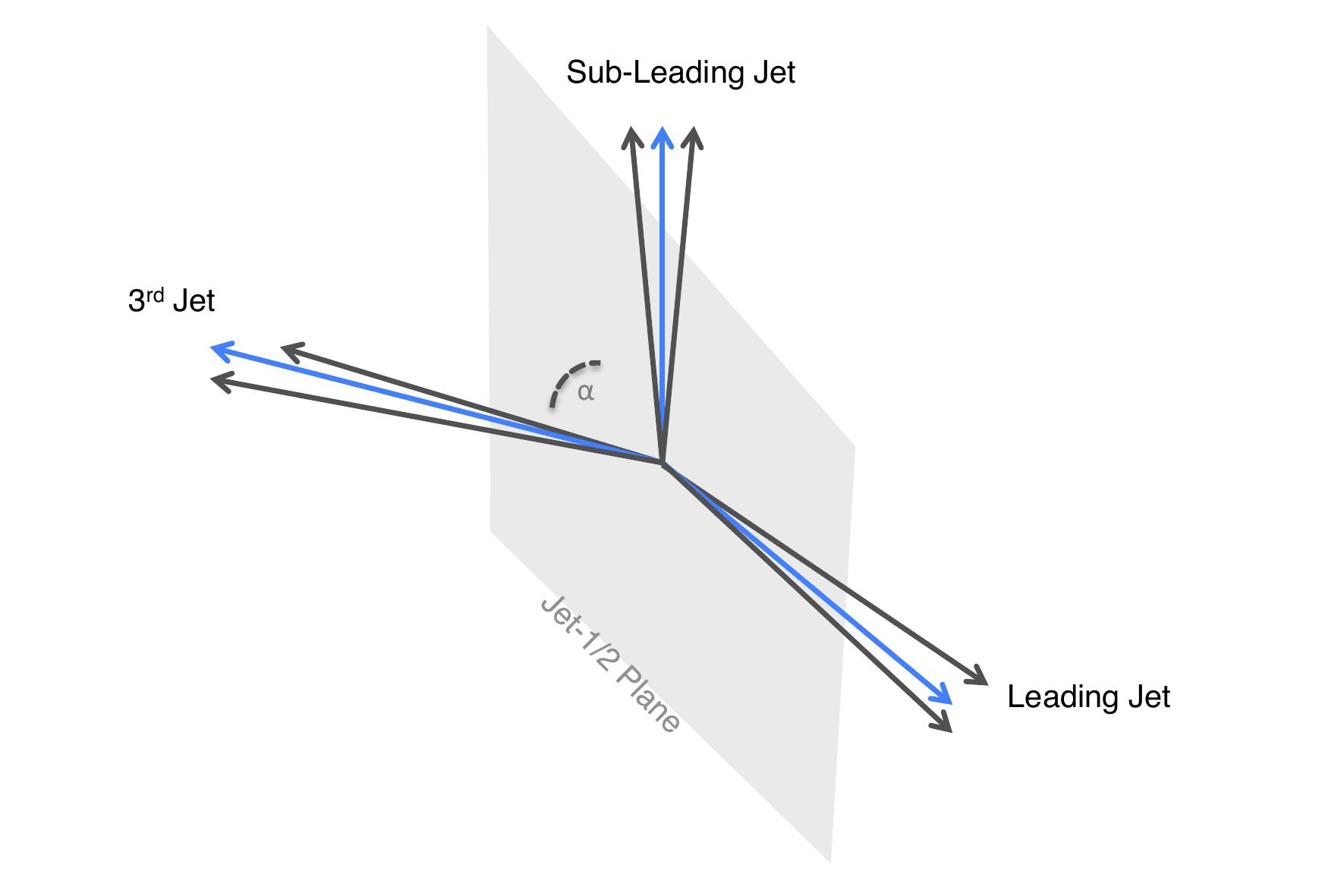}
\caption{Illustration of the definition of $\alpha$ in a three-jet system, where the leading jet is defined as the jet with the highest $p_T$. \label{fig:Alpha} }
\end{minipage}%
\hspace{0.1cm}
\begin{minipage}{7.3cm}
\centering
\includegraphics[width=0.93\linewidth]{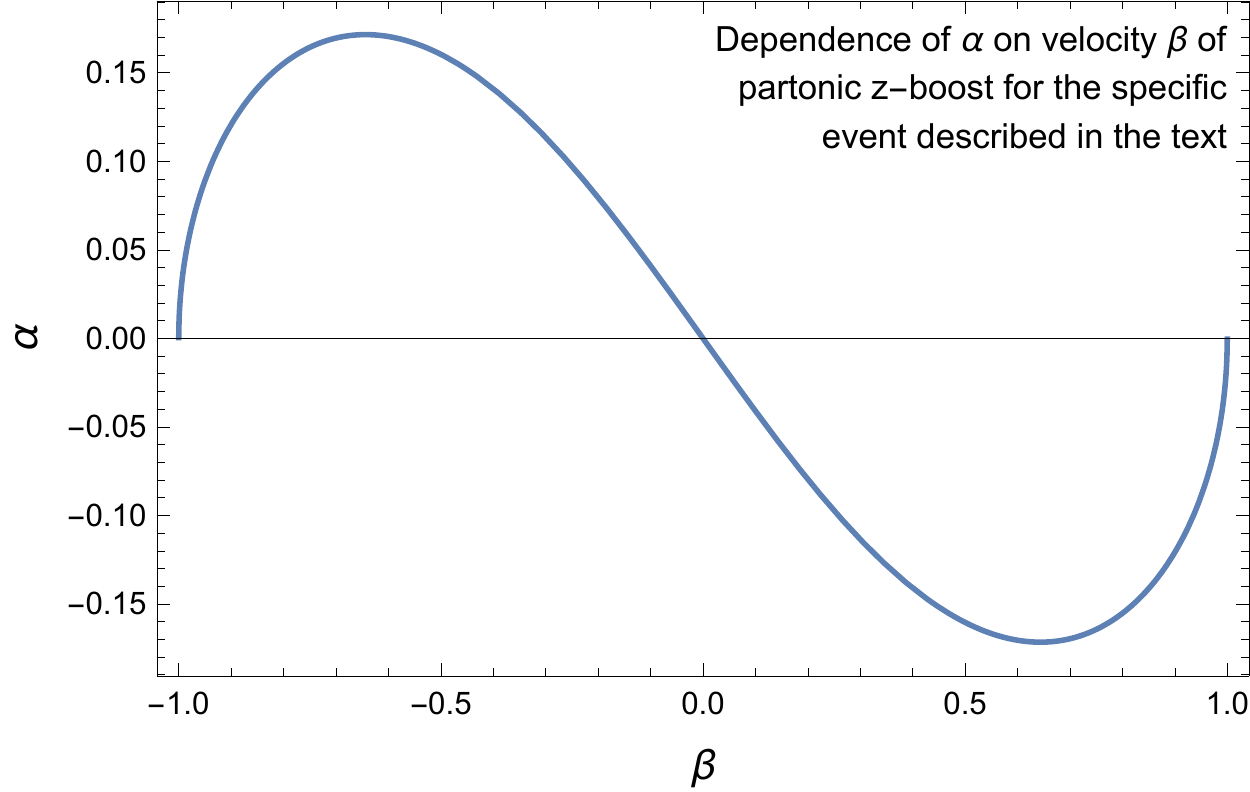}
\caption{Dependence of $\alpha$ on the longitudinal velocity $\beta$ of the three-jet system for the example configuration described in the text. \label{fig:BetaDependence} }
\end{minipage}
\end{figure}

A simple way of constructing a parity-odd variable from jets which is insensitive to rotations is to use a pseudoscalar. This simplest of these is arguably the scalar triple product $[\moma,\momb,\momc] \equiv (\moma \times \momb) \cdot \momc$, computed from the three jet three-momenta $\moma$, $\momb$ and $\momc$ ordered by the requirement \begin{align}p_T(\namea)\ge p_T(\nameb) \ge p_T(\namec)\label{eq:ptordering}.\end{align}Though the scalar triple product has the beneficial property of changing sign under parity, $(\moma, \momb, \momc) \rightarrow (-\moma,- \momb,-\momc)$, it has the disbenefit that it varies not only with the directions but also linearly with the magnitudes of each of the momenta.  We therefore choose instead to use a normalised variant whose dependence is primarily on the directions of the momenta:\footnote{The quantity $\sin\alpha$ does not depend {\it exclusively} on the directions of $\moma$, $\momb$ and $\momc$ because (\ref{eq:ptordering}) uses other properties of $\moma$, $\momb$ and $\momc$ (specifically $p_T$ ordering) to define which of them is which.  Nonetheless, the normalisation choice makes $\alpha$ completely independent of measurement errors on momentum magnitudes \textit{until} the point at which any measurement errors become large enough to change the $p_T$ ordering. Had the normalisation been achieved  by dividing by $|\moma||\momb||\momc|$ instead of by $|\moma \times \momb||\momc|$, the same insensitivity to small changes in momenta magnitudes would still be present, but the resulting variable would no longer have a simple geometric interpretation as the sine of an angle between two planes.  We opt for the former to preserve a geometric interpretation of the variable. We re-emphasise that there is nonetheless considerable arbitrariness in our choice of $\alpha$ -- which itself is attributable to the lacking concrete models predicting visible parity violation at the LHC.  Were such models to exist, we could speak quantitatively of `better' or `worse' choices of variable, and make optimisations. At the present time, however, we can only described the consequences and limitations of the choice we have made.  The body text provides a number of examples of alternative choices which, if made, would result in variables with very different regions of sensitivity.  We hope these ideas and others will be the subject of future investigations.} \begin{align}
   \frac{\moma \times \momb}{|\moma \times \momb|}
    \cdot \frac{\momc}{ |\momc|} = \sin\alpha.
\end{align}
We have denoted this quantity `$\sin \alpha$' since  it may be interpreted as the sine of the angle $\alpha$ which momentum $\momc$ makes with the plane spanned by $\moma$ and $\momb$, as shown in Figure~\ref{fig:Alpha}.  Positive angles $\alpha$ occur if and only if $\momc$ is on the same side of that plane as $\moma \times \momb$.

Note that magnitude of $\sin\alpha$ is not invariant under longitudinal boosts of the system of three jets.  
Even the sign of $\alpha$ can be altered by longitudinal boosts for certain jet momenta.  
Consider, for example, a co-planar set of jet momenta such as $\moma=(3,0,0)$, $\momb=(0,2,0)$ and $\momc=(1,1,0)$. 
As these are co-planar they have $\alpha=0$. 
If those momenta are then given a longitudinal boost, the value of $\sin\alpha$ ceases to be zero, but boosts in one direction generate equal and opposite values of $\alpha$ to boosts in the other direction.\footnote{Concretely, longitudinal boosts having velocities $\beta_z=\pm  1 / {\sqrt 2}$ change the momenta in the supplied example to   $\moma=(3,0,\pm 3)$, $\momb=(0,2,\pm 2)$ and $\momc=(1,1,\pm \sqrt 2)$ which have values of $\sin \alpha$ of  $\pm(- 1 /{\sqrt 3} + 1/ {\sqrt 6})$ respectively.  A plot showing how $\alpha$ in the above example varies for any $\beta$ is shown in Figure~\ref{fig:BetaDependence}.}
Consequently, any search for a new physics process which generates an $\alpha$-asymmetry in the frame of the hard process must contend with the possibility that such an asymmetry might be partly `washed away' by the longitudinal boosts which relate the hard-process frame (HPF) to the lab frame (LF).  
Evidently, the sort of processes most easily washed away would be those which have three co-planar (or almost co-planar) final-state momenta in the HPF and are generated from initial states (such as gluon-gluon or quark-quark) whose hard processes have symmetrically distributed longitudinal boosts.

Given that a large fraction of the total LHC cross section comes from gluon-gluon and given that conservation of momentum would require co-planarity in the HPF of the outgoing partonic momenta in any $2\rightarrow 2$ or $2\rightarrow 3$ parton-level process, one might be concerned that $\alpha$ appears to lack sensitivity to parity-violating effects in some important places!

Such concern is, however, premature.  The \textit{only} way in which a set of co-planar momenta can have a notion of handedness akin to parity is if the plane in which they live is orientable.\footnote{By orientable, we mean that it has a `front' which is distinguishable from its `back'.}   
It is for the same reason that the (planar) letter \textbf{R} can be distinguished from its mirror image \reflectbox{\textbf{R}} only if \textit{one} side of the paper on which both are printed is deemed `special'.  The plane separating two \textit{identical} interacting initial-state partons in the HPF is necessarily \textit{non}-orientable.\footnote{\ldots~assuming isotropy of the laws of physics.}
Therefore, although 
$2\rightarrow 2$ and
$2\rightarrow 3$ processes using 
gluon-gluon and quark-quark initial states are those in which one might have feared the aforementioned `wash out', these same initial states are necessarily those which can carry through no notion of handedness from the intial to the final state in the HPF anyway, let alone one that could be washed away by symmetric boosts from that frame to the LF
\footnote{Another way of explaining the same point is to consider the following question: 
``Can a $g g \rightarrow j j j$ new-physics processes result at the LHC in  
more `right-handed' events of the form `$\alpha_+$' with momenta
$
\{
\moma = 
(3,0,3), \momb=
(0,2,2), \momc=
(-1,-1,\sqrt 2) \}$
than `left-handed' events of the form `$\alpha_-$' with  momenta 
$
\{ \moma = 
(0,3,3), \momb=
(2,0,2), \momc=
(-1,-1,\sqrt 2) \}$ ?'' [The sets of momenta $\alpha_+$ and $\alpha_-$ map into each other under the exchange $x\leftrightarrow y$.] In answering this question, we would first note that both of these jet configurations are co-planar in the HPF as they have  $z$-components which are zero there.  Secondly, we could observe that in the HPF an event of type $\alpha_+$ can always be mapped into an event of type $\alpha_-$ by a 180 degree rotation which leaves the gluon-gluon initial state invariant. Therefore, if the equivalence principle holds and the laws of physics are isotropic, then events of type $\alpha_+$ should occur as frequently as events of type $\alpha_-$ in gluon-gluon interactions, and so the answer to the posed question is `No!'. If instead the laws of physics are not Lorentz invariant, or not isotropic, then the answer could instead be `Possibly!'.   A similar argument would fail if the initial state were asymmetric (such as quark-gluon) since then the 180-degree rotation would not leave the initial state invariant and it would not be possible to conclude that the number of $\alpha_+$ and $\alpha_-$ events need be identical.
}.

In contrast, asymmetric initial states (such as quark-gluon) \textit{are} capable of distinguishing forward and backward directions in the HPF, and so \textit{are} (at least in principle) able to generate detectable lab-frame asymmetries in $\alpha$.
It is important to realise that such asymmetries could be seen even for Lorentz- and rotationally-invariant $2\rightarrow 3$ new-physics processes having final state momenta which are co-planar in the HPF.  It is differences between the forward and backward boosts from parton distribution functions which would drive such an asymmetry, \textit{exploiting} the dependence of $\alpha$ on longitudinal boosts. Given the asymmetry is generated by this  mechanism, there is some motivation for biasing the signal region to events in which the three-jet system has a high absolute rapidity. 

In summary: the three-jet variable which we have called $\alpha$ can (in principle) detect parity-violating effects in a broad class of new-physics models at the LHC.  As the particular variable we used to illustrate our search is just one of many that could have been chosen, we make no claims of optimality -- indeed we have highlighted many situations in which it has no sensitivity to anomalous parity violation whatsoever. Nonetheless, we have: (i) demonstrated that parity violating models to which our variable \textit{is} sensitive do not need to abandon the equivalence principle, Lorentz invariance or isotropy, and (ii) noted that since $\alpha$ is defined in terms of only three objects the parity violating processes which $\alpha$ is sensitive to would probably include only those which couple to asymmetric initial states, such as quark-gluon.

In future work, it may  be interesting to consider other event variables which depend on \textit{two} or \textit{four} (rather than three) final state momenta.  A simple variable of the first sort might be the product $\Delta(\phi) \Delta(\eta)$ of the azimuthal and pseudorapidity differences between any two final state objects.\footnote{This amazing variable is invariant under longitudinal boosts and does not need the final state particles to be distinguishable or carry an ordering!}  Variables acting on four objects could show sensitivity to non-standard parity violating effects which do not need the initial state to provide an reference direction.  One example would be the four-jet (longitudinal boost invariant) event-variable $\delta$ defined by:
\begin{align}
	\delta \equiv \frac { 
	p^\mu(\namea)
	p^\nu(\nameb)
	p^\sigma(\namec)
	p^\tau(\named)
	\epsilon_{\mu\nu\sigma\tau} 
	} {
		p_T(\namea) 
		p_T(\nameb) 
		p_T(\namec) 
		p_T(\named) 
		}.
		\label{eq:delta}
\end{align}
The numerator of $\delta$ is the archetypical Lorentz-invariant pseudoscalar, so is certainly invariant with respect to longitudinal boosts and capable of demonstrating non-standard parity violating effects. The numerator makes $\delta$ dimensionless, and is also invariant under longitudinal boosts because it is built exclusively from transverse quantities.  A disadvantage of $\delta$, however, is that it requires the inputs to have been ordered in some way, as in (\ref{eq:ptordering}).  It is by no means a requirement that non-standard parity violating effects should come together with an ordering requirement on momenta.  It may also be useful, therefore, to consider completely symmetric pseudoscalars.  These are pseudoscalars which (unlike $\delta$) do not change their value if their input momenta are re-ordered. Examples include the four completely symmetric pseudoscalars  $S_1$, $S_2$, $S_3$ and $S_4$ defined by
\begin{align}
S_1(p(1),p(2),p(3),p(4)) &= 
   \sum_{i,j,k,l=1}^3 
    [i,j,k,l]
   (i \cdot j)
   (i \cdot j)
   (i \cdot k)
   ,
\\
S_2(p(1),p(2),p(3),p(4)) &= 
   \sum_{i,j,k,l=1}^3 
    [i,j,k,l]
   (i\cdot j)
(i \cdot k)
   (j \cdot l)
     ,
\\
S_3(p(1),p(2),p(3),p(4)) &= 
   \sum_{i,j,k,l=1}^3 
    [i,j,k,l]
 (i \cdot i)
(j \cdot j)
(i \cdot k)
   \qquad\text{and}
\\
S_4(p(1),p(2),p(3),p(4)) &= 
   \sum_{i,j,k,l=1}^3 
    [i,j,k,l]
 (i \cdot i)
(j \cdot k)
(i \cdot j),
\label{eq:sfour}
\end{align}
in which $(a\cdot b)$ stands for $p^\mu(a)p_\mu(b)$ and $[a,b,c,d]$ stands for $\epsilon_{\mu\nu\sigma\tau} p^\mu(a) p^\nu(b) p^\sigma(c) p^\tau(d)$.  $S_1$ to $S_4$ are not dimensionless, but they could easily be made so in the same way as $\delta$ in (\ref{eq:delta}).\footnote{Note that $S_3$ and $S_4$ are identically zero if evaluated on four-momenta belonging to massless particles.   $S_1$ and $S_2$ are therefore more relevance to jet-based variables than $S_3$ and $S_4$.}

\section{The CMS Detector and Reconstructed Objects,\label{Sec:CMS}}

The data used in this analysis have been recorded with the CMS detector at the LHC in the year 2012. CMS is a high-energy physics experiment which uses a superconducting solenoid, of 6 m internal diameter, with a magnetic field of 3.8 Tesla. The inner detector (ID) of CMS can reconstruct trajectories of charged particles using a silicon pixel and strip tracker. Electrons and photons are identified and measured in a crystal electromagnetic calorimeter (ECAL), while energies of hadrons or hadronic particle jets are determined in a brass/scintillator hadron calorimeter (HCAL). Muons are identified and measured in the muon system (MS), based on gaseous detectors, which surround the hadronic calorimeter and are embedded in the steel flux-return yoke of the magnet system. CMS uses a right-handed coordinate system. Its origin is defined at the interaction point of the proton collisions, the $x$-axis is pointing towards the center of the LHC, the $y$-axis pointing upwards and the $z$-axis along the counterclockwise-beam direction. The polar angle $\theta$ is measured from the positive $z$-axis, however, mostly expressed in terms of the pseudorapidity $\eta$, defined by $\eta = - \ln(\tan(\theta/2))$. The azimuthal angle $\phi$ is measured in the $x$-$y$ plane. We refer to \cite{Chatrchyan:2008aa} for a detailed description of the CMS experiment.

The main objects used in this analysis are muons and particle jets, which are briefly discussed in the following. CMS employs a particle-flow algorithm that provides a complete description of the event and identifies electrons, muons, photons, charged hadrons, and neutral hadrons. Muons are reconstructed from a global fit of hits in the MS and the ID, seeded by tracks in the muon system \cite{CMS-PAS-MUO-10-002}. In this analysis, we require each muon to have a minimal transverse momentum of $p_T>25$ GeV within a pseudo-rapidity range of $|\eta|<2.1$, corresponding to the single muon trigger coverage. In addition, standard quality cuts on the number of hits in the ID and the MS, the $\chi^2$ of the fit as well as on the impact parameters are applied, also following previous CMS measurements \cite{Chatrchyan:2014mua}. 

Hadronic jets are reconstructed using an anti-$k_t$ algorithm \cite{Cacciari:2008gp, Cacciari:2011ma} with a radius parameter of 0.5 based on particle-flow objects \cite{CMS-PAS-PFT-10-001, CMS-PAS-PFT-09-001, Sirunyan:2017ulk}, where the clustering algorithms rejects objects that are coming from a pile-up vertex. A jet area method is used to correct for the remaining pile-up contributions \cite{Cacciari:2007fd}. Since the four-momenta of particle-flow objects is summed, the jets can be massive, In this analyses, we only study jets with a minimal transverse momentum of $p_T>30$ GeV and a jet rapidity of $|y|<2.4$, since this region allows for the best jet resolution and pile-up rejection. In addition, certain quality criteria on the reconstructed jet properties, such as energy fraction in the ECAL and HCAL or the number of particle-flow objects is applied, following the standard CMS recommendations. Moreover, jets are required to have a minimal distance of $\Delta R>0.5$ to all reconstructed electrons, muons and photons candidates. The reconstructed transverse momenta of jets is used to order them as the $1^{st}$, $2^{nd}$ and $3^{rd}$ jet according to their $p_T$ and denoted as $j_1$, $j_2$ and $j_3$, respectively.

This analysis is built around the CMS Open Data Software Framework available at \cite{CMS:OpenData}.  Version \texttt{CMSSW\_5\_3\_32} \cite{CMS:5:3:32} was used. A dedicated open-source framework, \textsc{Bacon} \cite{cit:bacon}, which was used for several published studies of the CMS Collaboration, e.g.~\cite{Chatrchyan:2014mua}, is used to read the Analysis Object Data (AOD), extracting information on reconstructed objects as well as generator level data, if available.
The \textsc{Bacon} software framework is also used to apply a GoodRun-List selection provided by the CMS Open Data project \cite{CMS:GoodRunList}, as well as calibration constants for particle jets as well as \MET~observables, leading to a separate output-format based on \textsc{Root}-tree objects. The typical event size of one simulated top-quark pair event in the \textsc{Bacon}-output format amounts to 5 kB. 
For this work, we used an additional software package, which reduces the output files of \textsc{Bacon} further and transforms them into a plain \textsc{Root}-NTuple, denoted as \textsc{ODNTuple} in the following with an average event size of 0.8 kB. 
Our analysis is based on these \textsc{ODNTuple} data. The full software, which has been used for this work, was previously used and validated for \cite{Apyan:2019ybx}.

\section{Selected Open Data and Simulated Monte Carlo Samples}

The data acquisition system of CMS records only the event information of collisions with dedicated signatures due to the high-collision rate and the limited bandwidth for data-processing. The data used in this analysis has been collected when significant hadronic activity was present in the event, in particular if the \texttt{HLT\_PFJet320} trigger had fired, targeted to select events with at least one reconstructed jet with a transverse energy of 320 GeV.  In total, files corresponding to 289~\ipb\ from the CMS Open Data \cite{CMS:QCDData} have been processed, of which 166~\ipb\ fired the  \texttt{HLT\_PFJet320} trigger (Table \ref{tab:DataSamples}). The study of a larger data-set was beyond the available computing resources. The integrated luminosity has been calculated using the public available GoodRun-List as well as the corresponding tools. In total, 831,662 triggered events, which pass the GoodRun-List requirement, are used in this analysis.

\begin{table}[tb]
\footnotesize
\begin{center}
\begin{tabularx}{\textwidth}{l | l | c}
\hline
Data stream / Trigger name							& Dataset Name										&	$\int L dt $ \quad$[\ipb]$	\\
\hline
Incl. Jet-Triggers / \texttt{Trigger\_HLT\_PFJet320}				& JetHT\_Run2012C \cite{CMS:QCDData}						&	289	/ 166			\\
\hline
\end{tabularx}
\caption{Overview of data samples used in this analysis together with the corresponding integrated luminosity and the triggers, which have been used during the data taking.\label{tab:DataSamples}}
\end{center}
\end{table}

An overview of the various signal and background Monte Carlo (MC) samples used in this analysis is given in Table \ref{tab:MCSamples}, indicating the underlying physics process, the dataset name and the corresponding inclusive cross-section. The inclusive QCD MC was generated with \textsc{Pythia6}~\cite{Sjostrand:2006za} and covers a jet transverse momentum range between 50 and 3000 GeV. In addition, a 4-jet sample is used as alternative QCD sample, which is based on the \textsc{Alpgen} generator \cite{Mangano:2002ea}. The Drell-Yan processes (W/Z) in the muon decay channel were generated using with the \textsc{PowhegBox} Monte Carlo program~\cite{Alioli:2008gx, Alioli:2010xd} interfaced to the \textsc{Pythia} v.6.4.26 ~\cite{Sjostrand:2006za}  parton shower model. All other processes are modeled with the tree-level matrix element event generator \textsc{MadGraph} v5.1.3.30 \cite{Alwall:2011uj} interfaced with \textsc{Pythia} v6.4.26. In all samples the CT10 PDF set \cite{Gao:2013xoa} and the Z2* \textsc{Pythia}6 tune ~\cite{Chatrchyan:2013gfi,Khachatryan:2015pea} are used. The decay of tau-leptons is modeled using the \textsc{Tauola} program \cite{Davidson:2010rw}, while the emission of photons from finale state leptons uses  \textsc{Pythia}6.  The strong coupling constant $\alpha_s$ has been set to $0.130$ at the $Z$-boson mass scale for all matrix element calculations. The effect of multiple interactions per bunch crossing (pile-up) has been simulated by overlaying MC-generated minimum bias events. The \textsc{Geant4} program was used to simulate the passage of particles through the CMS detector \cite{Agostinelli:2002hh}. 

The simulated event samples are reweighted to describe the distribution of the number of pile-up conditions in the data by reweighting the $\rho$ parameter distribution, where $\rho$ denotes the diffuse offset energy density~\cite{Khachatryan:2016kdb}. Moreover, a reweighting of the  longitudinal position of the primary  $pp$ collision vertex of the MC samples to data has been performed.


\begin{table}[tb]
\footnotesize
\begin{center}
\begin{tabularx}{\textwidth}{l | l | l}
\hline
Process											& Dataset Name												&	$\sigma$ [pb]		\\
\hline
QCD	 (Inclusive, \textsc{Pythia6})						& QCD\_Pt-15to3000\_TuneZ2star \cite{CMS:QCD1}				& $2\cdot10^{10}$\\
QCD	 (4-Jets , \textsc{Alpgen})							& QCD4Jets\_Pt-400to5600\_TuneZ2Star	\cite{CMS:QCD2}			& - \\ 
\hline
$pp \rightarrow t\bar{t}+X \rightarrow 2l2\nu2b+ X$				& TTJets\_FullLeptMGDecays\_TuneP11TeV\_8TeV	\cite{CMS:MCTTLep}	&	112.3				\\
$pp \rightarrow t\bar{t}+X \rightarrow 1l1\nu2q2b+ X$			& TTJets\_SemiLeptMGDecays\_8TeV \cite{CMS:MCTTSemi}				&	107.2				\\
$pp \rightarrow t\bar{t}+X \rightarrow 4q2b+ X$					& TTJets\_HadronicMGDecays\_TuneP11mpiHi\_8TeV \cite{CMS:MCTTHad}	&	25.8					\\
$pp \rightarrow WW+X \rightarrow 2l2\nu+ X$					& WWJetsTo2L2Nu\_TuneZ2star\_8TeV \cite{CMS:MCWW}				&	5.8					\\
$pp \rightarrow WZ+X \rightarrow 3l1\nu+ X$					& WZJetsTo3LNu\_8TeV\_TuneZ2Star \cite{CMS:MCWZ}					&	1.1					\\
$pp \rightarrow Z/\gamma^*+X \rightarrow \mu^+\mu^- + X$		& DYToMuMu\_M-20\_CT10\_TuneZ2star\_v2\_8TeV \cite{CMS:MCZmumu}	&	1931					\\
\hline
\end{tabularx}
\caption{Overview of data samples and simulated event samples used in this analysis together with the corresponding inclusive cross-sections. Leptonic decay ($e,\mu,\tau$) are denoted with $l$.\label{tab:MCSamples}}
\end{center}
\end{table}


\section{\label{Sec:Calibration}Calibration}

Even though the full detector simulation of CMS provides a very good description of the expected event signatures, some remaining differences in reconstruction, trigger and isolation efficiencies as well as in the momentum and energy scales and resolutions between MC simulation and data are present. Dedicated corrections are applied to minimize these differences for the relevant objects in this analysis and are discussed in the following. 

The official CMS calibration and corrections for particles jets, in particular the jet energy scale (JES) and the jet energy resolution (JER), has been applied within the \textsc{Bacon} framework. These jet corrections and uncertainties were derived from the simulation, and are confirmed with in situ measurements using the energy balance of dijet and photon+jet events \cite{Khachatryan:2016kdb}. A reduced set of systematic variations is used to estimate JES and JER uncertainties on the final measurement. In particular, the JES is variated by 2\% for $y^{jet}<1.3$ and by 3\% for $y^{jet}>1.3$, following \cite{Khachatryan:2016kdb}. The JER is varied by 20\% for $30<\ET^{jet}<100$ GeV, by 10\% for $100<\ET^{jet}<1$ TeV and by 5\% above.  The absolute angular $\phi$- and $\eta$-resolution for reconstructed jets with a transverse momentum of 50 GeV is between 0.01 and 0.02 and improves for larger \pT\ values. We assume a relative uncertainty of 50\% on the both angular resolutions.

\begin{figure}[tb]
\begin{center}
\includegraphics[width=7.3cm]{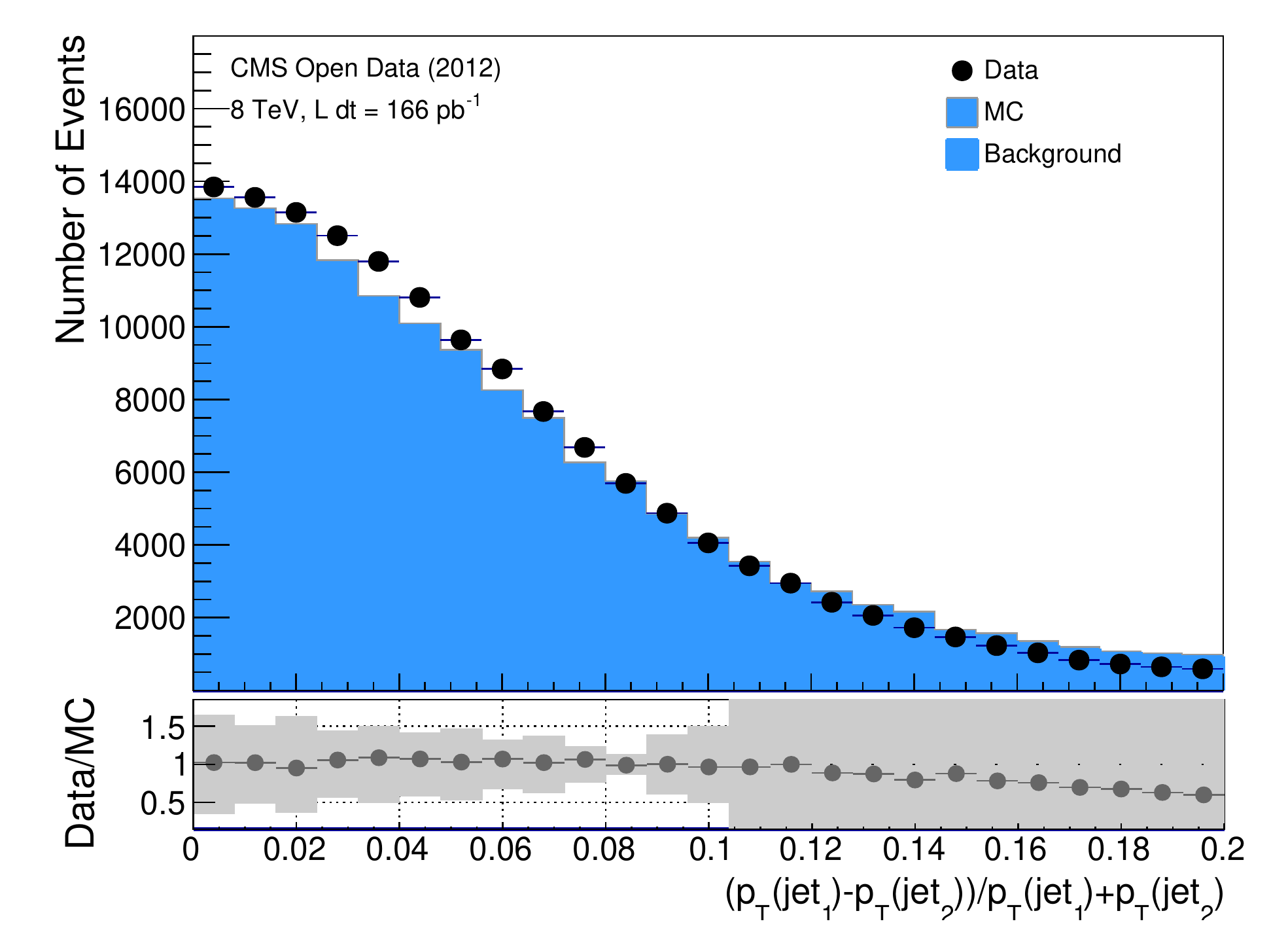} 
\hspace{0.1cm}
\includegraphics[width=7.3cm]{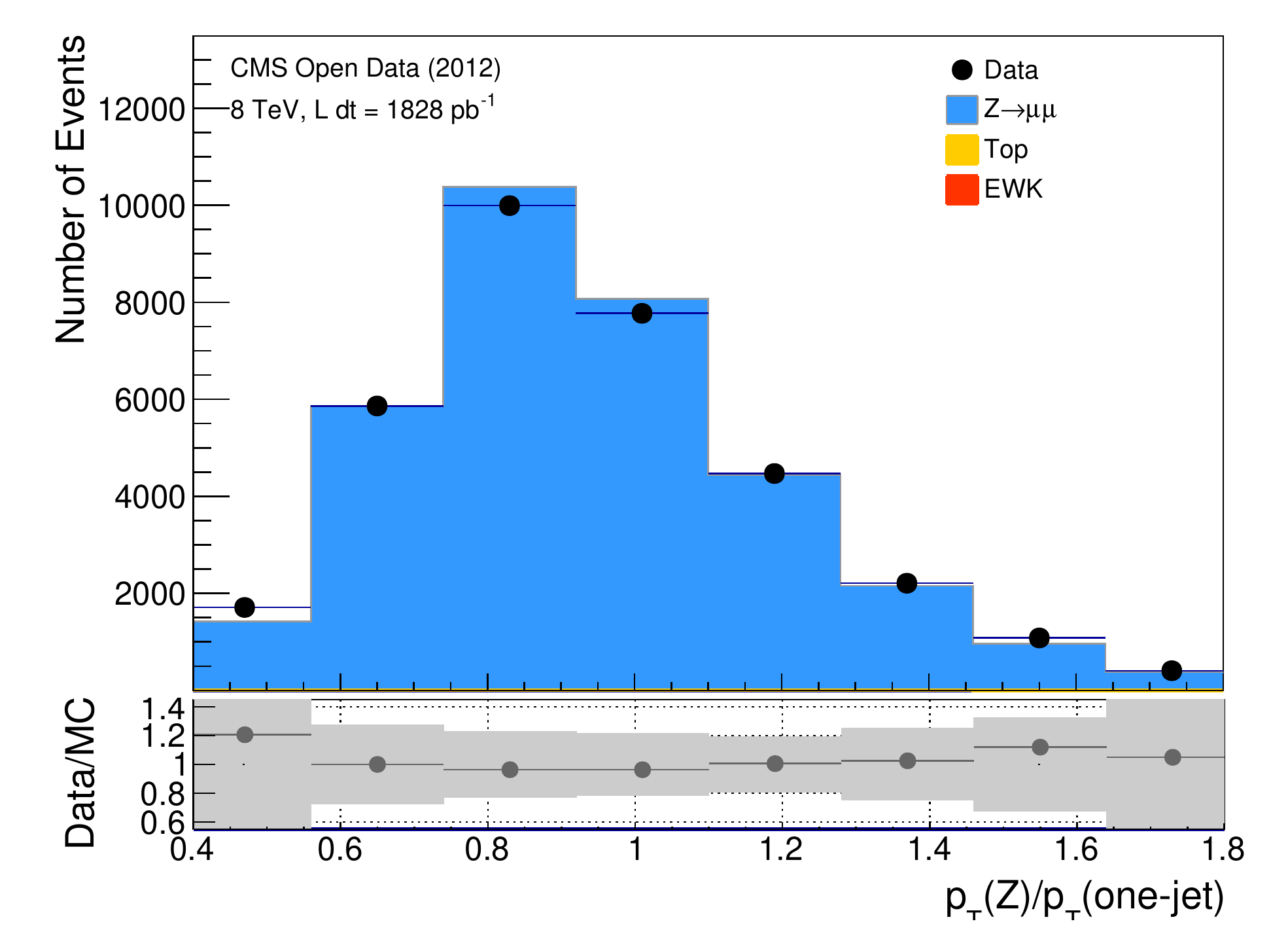}
\caption{\label{fig:JetPerformance} Left: Distribution of the $\DeltarelpT = (\pT^{j_{1}}-\pT^{j_{2}})/( \pT^{j_{1}}+\pT^{j_{2}})$ in Data and MC. Right: Comparison of the ratio of measured $\pT(Z)$ and the measured jet energy $\ET$ for $Z$-boson events in the muon decay channel with exactly one jet with $\ET>30$ GeV and $20<\pT(Z)<50$ GeV for Data and MC. All MC-corrections have been applied. The grey band indicates the corresponding systematic uncertainties.}
\end{center}
\end{figure}

Events containing exactly two jets with $\pT>50$~GeV within $y^{jet}<2.4$ are used to test the calibration as well as the assigned uncertainties. A comparison of the
$\DeltarelpT = ( \pT^{j_{1}}-\pT^{j_{2}})/( \pT^{j_{1}}+\pT^{j_{2}})$ distributions for data and MC, including the assigned systematic uncertainties is shown in Figure~\ref{fig:JetPerformance}, where agreement between data and MC can be seen.

To further validate the applied corrections, $Z$-boson candidate events with an associated production of at least one jet have been selected in data and compared to full simulated Monte Carlo samples. $Z\rightarrow \mu^+\mu^-$ candidate events are selected by requiring events with exactly two oppositely charged, isolated muons with a minimal $p_T$ of 25 GeV within $|\eta|<2.1$. 
Jets are required to have a minimal transverse momentum of $p_T>30$ GeV within a rapidity range of $|y|<2.4$. The invariant mass of these two lepton candidates has to be between 66 and 116 GeV. 
This selection ensures a nearly background free selection of $Z$-boson candidates. 
In order to test the jet calibration, $Z$-boson events with exactly one reconstructed jet are further divided into evens with $20<\pT(Z)<50$ GeV and $50<\pT(Z)<100$ GeV. 
The transverse momentum of the $Z$-boson, precisely measured by its decay leptons, should be balanced in a first approximation by the transverse energy of the select jet, hence the ratio of $\pT(Z)/\pT^{jet}$ should peak around 1. 
The comparison of data and MC of this ratio is also shown in Fig.~\ref{fig:JetPerformance}, where a good agreement within the assigned systematic uncertainties can be seen. 
This study has been repeated in \cite{Apyan:2019ybx} with higher jet multiplicities and higher values of $\pT(Z)$ in the final state, all indicating good closure.


\section{Analysis Strategy, Signal Selection and Background Estimation\label{sec:Analysis}}

It is the goal of this analysis to probe a possible asymmetry between positive and negative values of $\alpha$ (defined in Section~\ref{Sec:ParityViolation}) in events containing at least three jets in proton-proton collisions at a center of mass energy of $\sqrt{s}=8$ TeV.  The dependence of this asymmtry will be investigated in bins of the invariant di-jet mass of the leading and sub-leading jet, $m_{12}$. In addition, the normalized $\alpha$ distribution in data is measured and corrected for detector effects. 

Events with at least three jets with a minimal transverse momentum of $p_T>$ 50 GeV within a rapidity range of $|y|<2.4$ are selected for this analysis. Events with a fourth jet with $p_T>30$ GeV within $|y|<3.0$ are vetoed. Standard noise cleaning cuts as well as cuts to reject jets from pile-up collisions are applied. In addition, it is required that the minimal difference between the transverse momentum of the leading and the sub-leading jet be at least 30 GeV, i.e.~$\Delta p_T^{j_{1},j_{2}}=p_T^{j_{1}}-p_T^{j_{2}}>$ 30 GeV, and that the minimal transverse momentum difference between the sub-leading and sub-sub-leading be at least 50 GeV, i.e.~$\Delta p_T^{j_{2},j_{3}}=p_T^{j_{2}}-p_T^{j_{3}}>$ 50 GeV.  These requirements ensure that small changes to jet energies (e.g.~from reconstruction effects) do not lead to changes in the $p_T$ ordering of the jets, and so do not result in discrete changes in the orientation of the reference plane.
Moreover, we veto fully back-to-back events by requiring that the $\Delta\Phi$ value between $j_1$ and $j_2$ is smaller than 3.0. This selection reduces uncertainty in the normal of the reference plane, since that normal becomes undefined when $j_1$ and $j_2$ are parallel. In total 29,063 events pass this selection. The selected events are further divided along the invariant mass between the leading and sub-leading jet within the ranges $500<m_{12}<700$~GeV, $700<m_{12}<900$ GeV and $900<m_{12}<2000$~GeV as well as for two rapidity regions of the three jet system, $y_{jjj}$, with $|y_{jjj}|<0.7$ and $0.7<|y_{jjj}|$. 

Possible contributions to the selected final states involving electroweak decay processes, such as the production of $W+\text{jets}$ and $Z+\text{jets}$ in the hadronic decay mode or the production of $t \bar t$, are estimated by MC simulations and found to be below 0.5 \%, limited by the available MC statistics. As an additional cross-check, events with two jets with $p^{j_1}_T>350$ GeV and $p^{j_2}_T>100$ GeV and one additional reconstructed isolated muon with a transverse momentum of at least 50 GeV have been selected. Those events are expected to come from $W/Z$+jets, $t\bar t$ or $WW$ decays, with at least one vector boson decaying leptonically. In total, 31 events pass this selection, of which no events with two muons were found. Assuming that these muons all stem from leptonic $W$-boson decays, we expect $\approx$ 310 electroweak background events, i.e.~a background contribution of 1\%, which we take as an estimate in the following. 

Both QCD MC samples (Table \ref{tab:MCSamples}) have been used to study migration effects, i.e.~a change in the jet ordering or a wrong charge assignment of $\alpha$ using MC truth information. The minimal \pT\ difference between the jets as well as the $\Delta\Phi(j_1,j_2)$ requirement results in only 1.5\% of events, in which the jet-$\pT$ ordering is different between reconstruction and generator level and 1\% of the cases, which lead to wrong sign-assignment of $\alpha$. 

\begin{figure}[tb]
\begin{center}
\includegraphics[width=7.3cm]{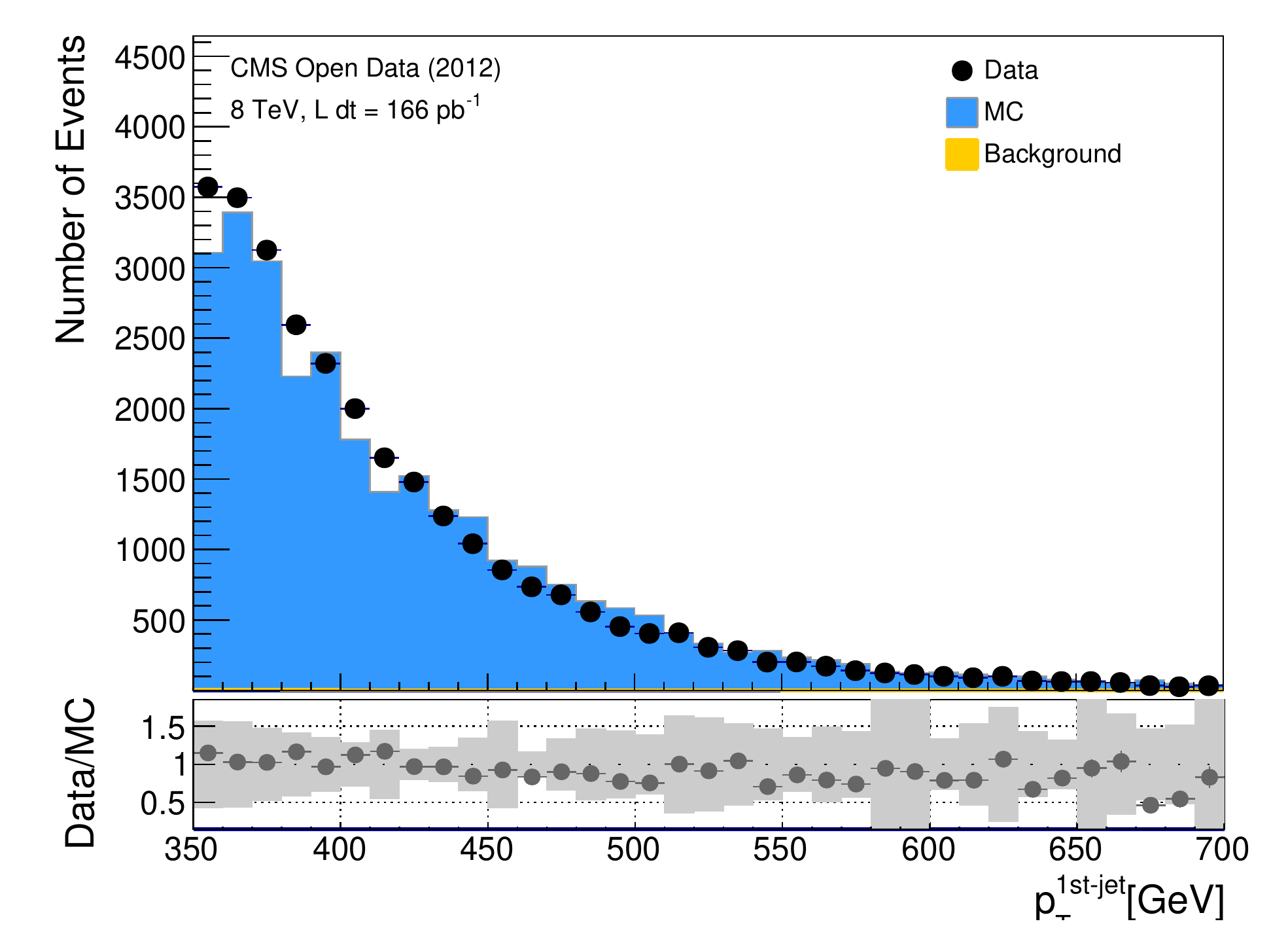} 
\hspace{0.1cm}
\includegraphics[width=7.3cm]{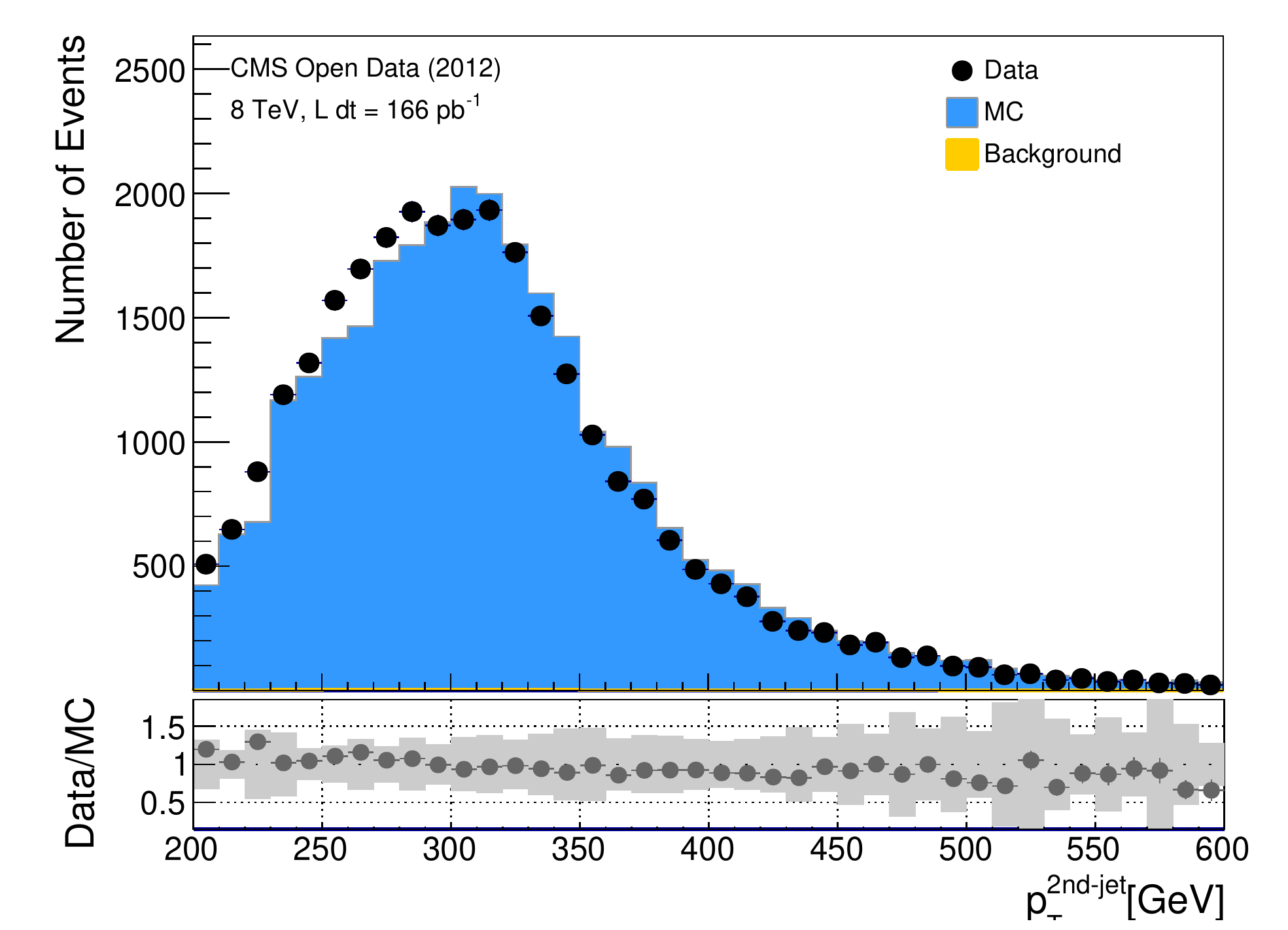}
\hspace{0.1cm}
\includegraphics[width=7.3cm]{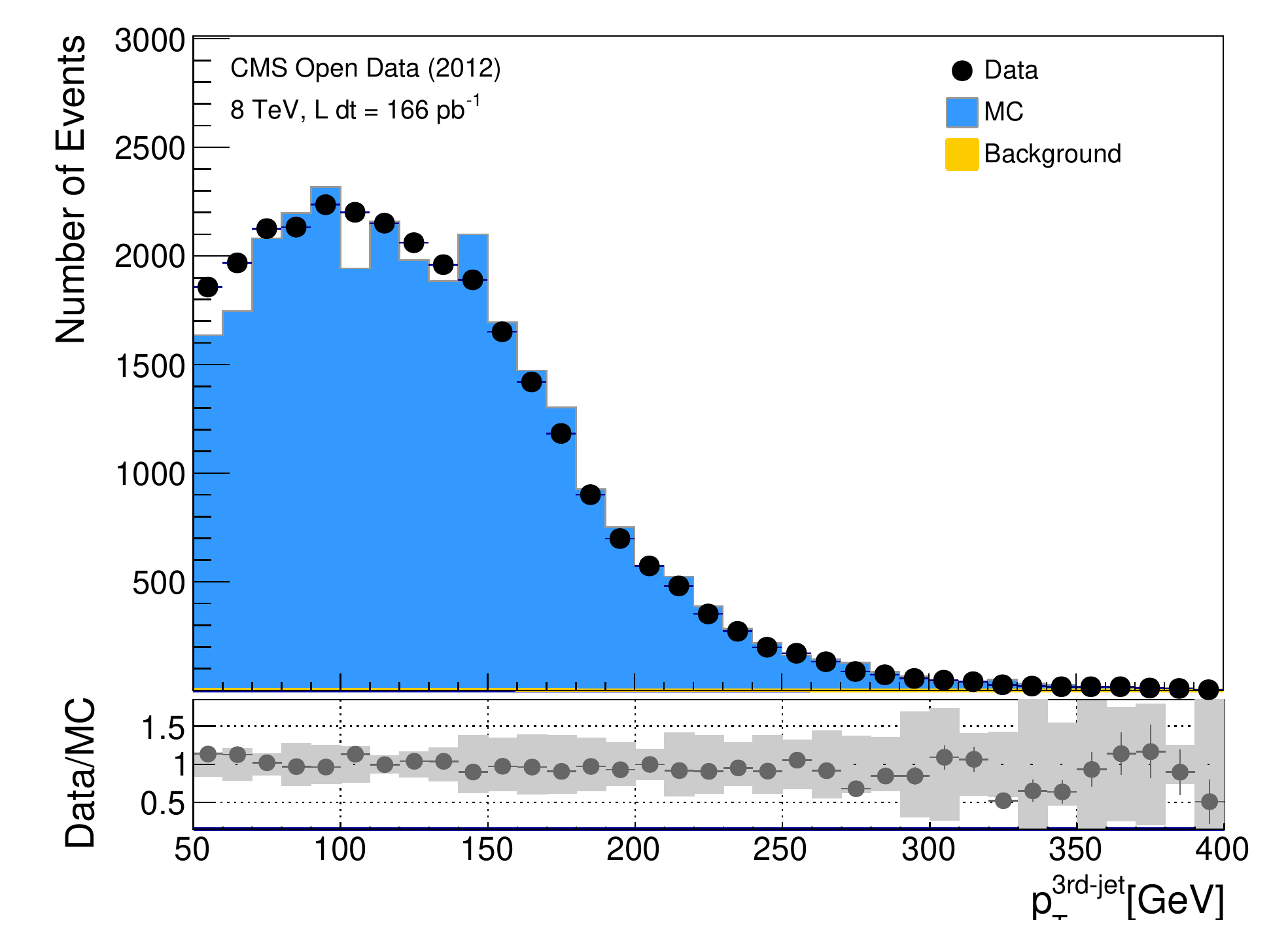}
\hspace{0.1cm}
\includegraphics[width=7.3cm]{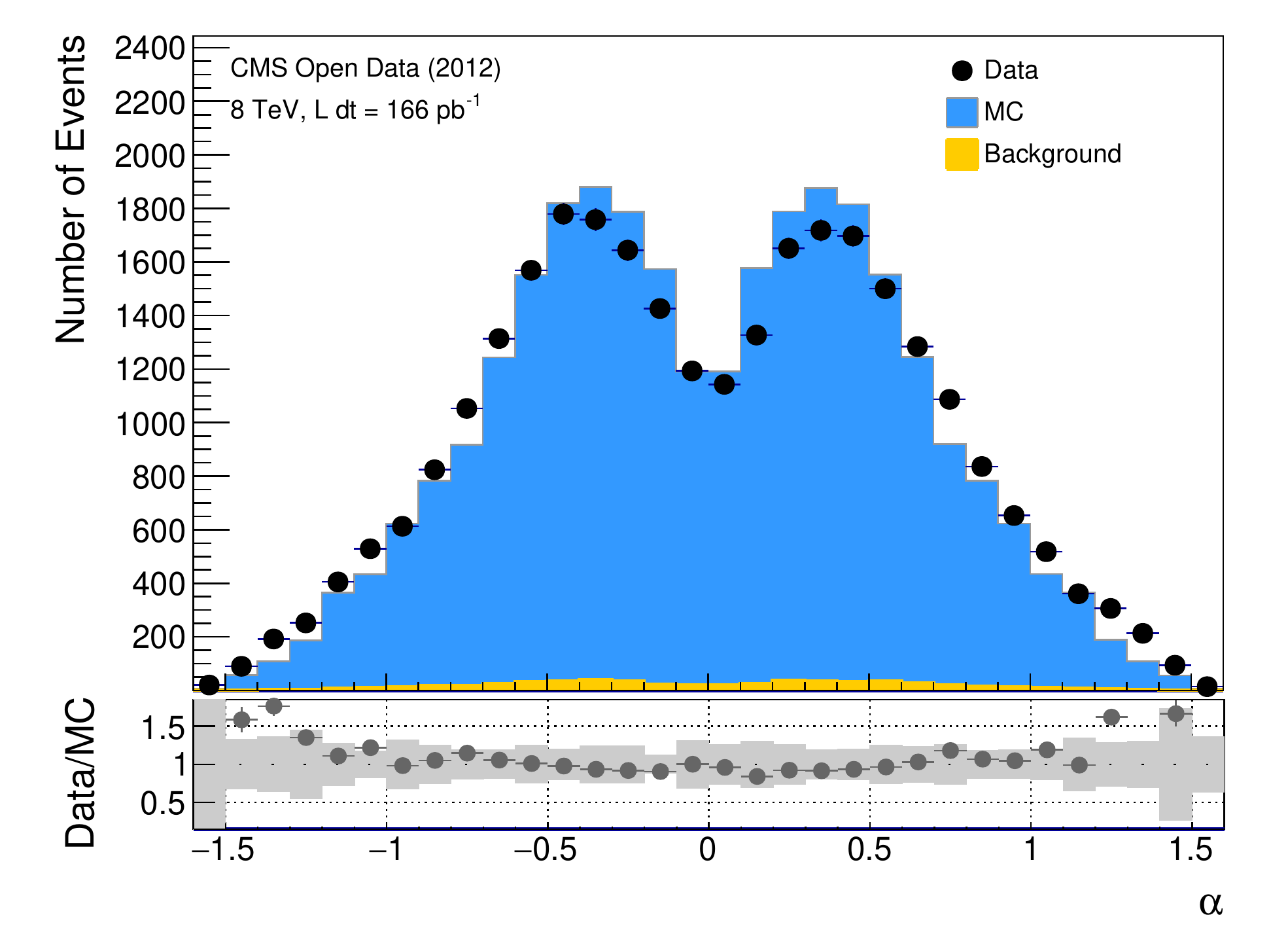}
\caption{\label{fig:DataMCReco} Comparison of the \pT\,\,distributions for the leading (upper left), sub-leading (upper right) and sub-sub-leading jets (lower left) for data and MC, as well as the inclusive $\alpha$ distribution (lower right) for data and MC. The grey band indicates the corresponding systematic uncertainties.}
\end{center}
\end{figure}

In order to improve the description of the MC samples with data, the transverse momenta distribution of the leading and sub-leading jets have been reweighted to data by a two-dimensional approach on an event-by-event basis using all events containing three jets with $p_T>$ 50 GeV within a rapidity range of $|y|<2.4$. The reweighting is based on two dimensional histograms with the $\pT$ of the leading and sub-leading jet on the axis, which is filled for data and MC. The actual event weights are derived by the ratio of the normalized data histogram over the corresponding normalized histogram of the MC set. It should be noted, that the fiducial volume which is used for the reweighting is significantly larger than the actual fiducial volume of the analysis. Figure \ref{fig:DataMCReco} shows the reconstructed $p_T$ distributions after the reweighting procedure of all three jets in all selected events as well as the inclusive distribution of $\alpha$ in data as well as MC simulations, where the distributions are normalized. Data and MC agree well within their uncertainties.

\section{Raw Asymmetry Measurements\label{sec:ResultsAndDatasetAsymmetry}}

The $\alpha$ asymmetry, $A_\alpha$, can be defined as the difference between the sum over the fiducial cross-section in each measurement bin $i$ of $\alpha$ for positive and negative regions of $\alpha$, normalized by the overall cross section, i.e.

\begin{equation}
A_\alpha = \frac{\sum_{i, \alpha>0} d\sigma_i/\sigma^{fid} - \sum_{i, \alpha>0} d\sigma_i/\sigma^{fid}}{\sum_{i, \alpha>0} d\sigma_i/\sigma^{fid} + \sum_{i, \alpha>0} d\sigma_i/\sigma^{fid}} = \frac{\sum_{i, \alpha>0} d\sigma_i/\sigma^{fid} - \sum_{i, \alpha>0} d\sigma_i/\sigma^{fid}}{\sigma}
\end{equation}

The asymmetry $A_\alpha$ can be approximated by a pure data driven quantity $A^{data}_\alpha$, that reduces to a pure counting exercise:
\begin{equation}
A^{data}_\alpha = \frac{N_{\alpha<0} - N_{\alpha>0}}{N_{\alpha<0} + N_{\alpha>0}}, 
\end{equation}
where $N$ is the number of events in a particular region of $\alpha$. In the following we assume that $A^{data}_\alpha$ is, to a good approximation, equal to $A_\alpha$, since detector related resolution and efficiency effects are expected to be symmetric in $\alpha$, i.e.~affect events with positive and negative values of $\alpha$ in the same way and hence cancel in the ratio. This was tested and confirmed using MC simulations. The resulting asymmetries $A_\alpha$ for the full inclusive selection, as well as for the three invariant mass regions of the leading and sub-leading jet, are shown in Table \ref{tab:AlphaResults}. The separation in different invariant di-jet masses allows to probe energy dependencies of parity violating effects. All asymmetries are compatible with the Standard Model expectation of zero, within the statistical uncertainties. 

\begin{table}[tb]
\footnotesize
\begin{center}
\begin{tabularx}{\textwidth}{l |c |c |c }
\hline
Selection		& Number of Events ($\alpha<0$)	& Number of Events ($\alpha>0$)	& Asymmetry $A_\alpha$		\\
\hline
$m_{jj}$ inclusive		& 14661			& 14402			& $0.009 \pm 0.006$	\\
\hline
500<$m_{jj}$<700 GeV	& 3211			& 3189			& $0.003 \pm 0.012$	\\
700<$m_{jj}$<900 GeV	& 5615			& 5483			& $0.011 \pm 0.009$	\\
900<$m_{jj}$<2000 GeV	& 5646			& 5551			& $0.008 \pm 0.009$	\\
\hline
$|y_{jjj}|$<0.7			& 10192			& 9993			& $0.010 \pm 0.007$	\\
0.7<$|y_{jjj}|$			& 4469			& 4409			& $0.007 \pm 0.011$	\\
\hline
\end{tabularx}
\caption{Measured asymmetry of $\alpha$ for the inclusive selection as well as three invariant mass regions of the leading and sub-leading jet as well as two rapidity regions of the three jet system $y_{jjj}$. The uncertainties are statistical only. \label{tab:AlphaResults}}
\end{center}
\end{table}

\begin{figure}[tb]
\begin{center}
\includegraphics[width=7.3cm]{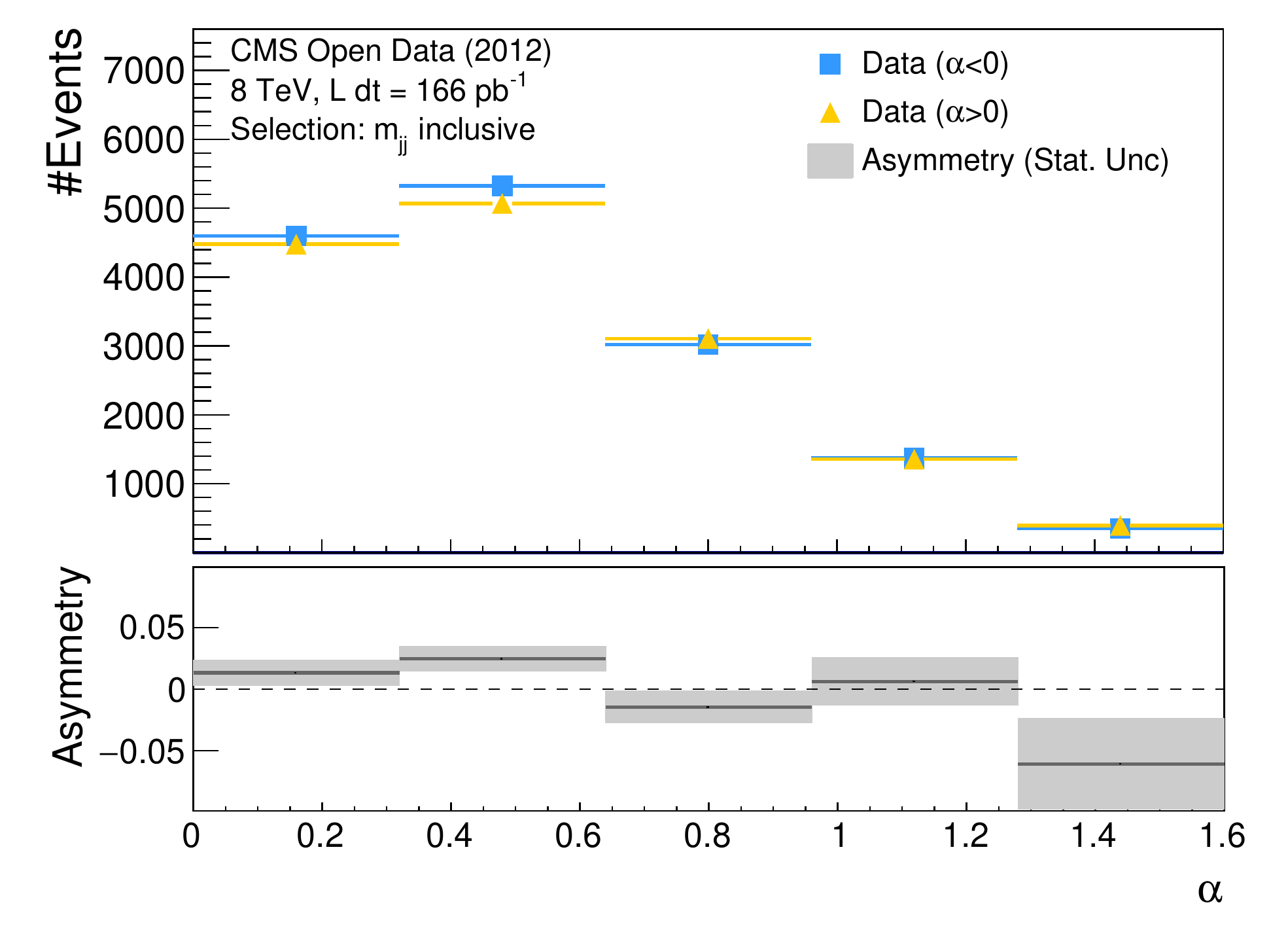} 
\hspace{0.1cm}
\includegraphics[width=7.3cm]{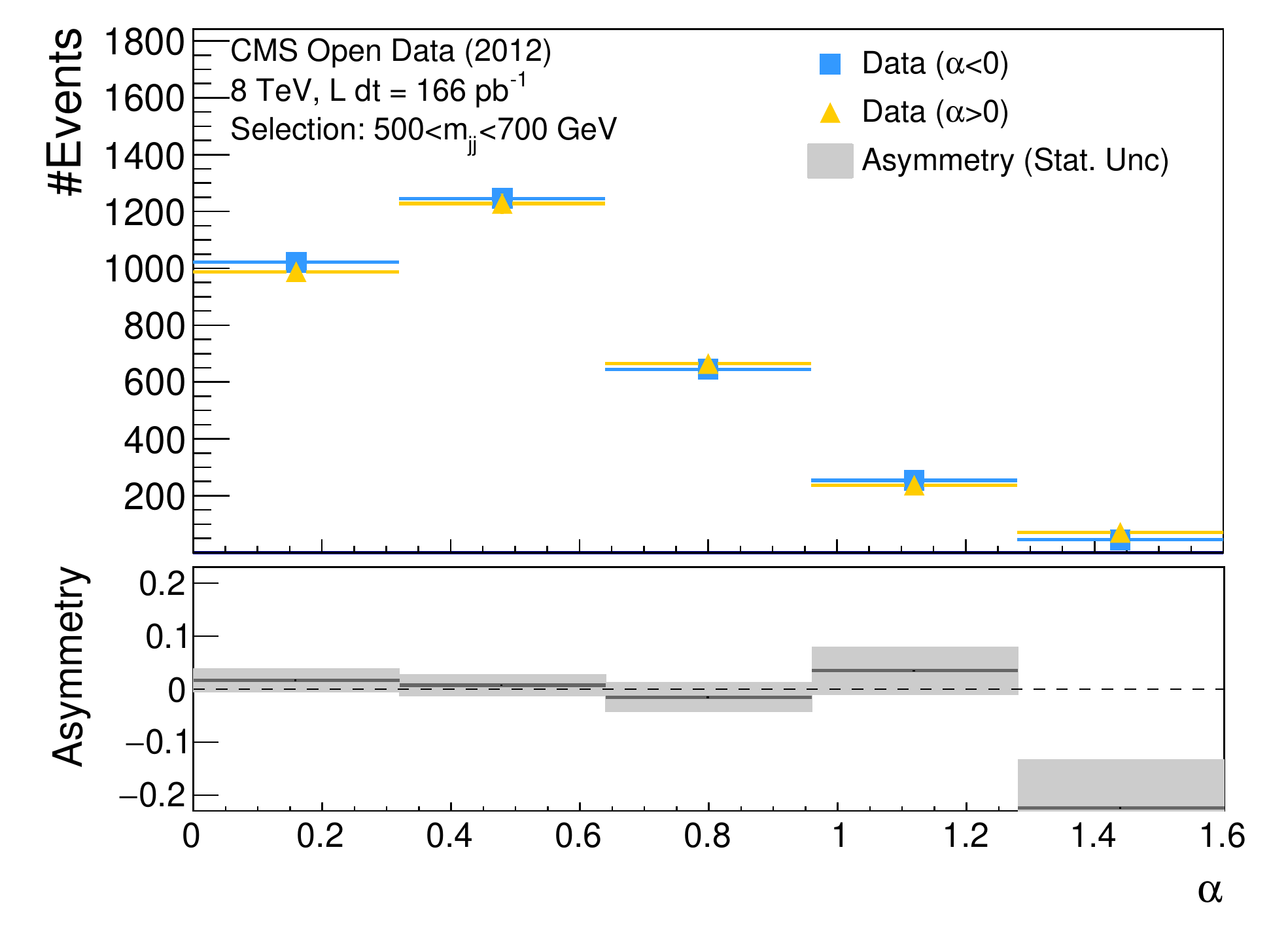}
\hspace{0.1cm}
\includegraphics[width=7.3cm]{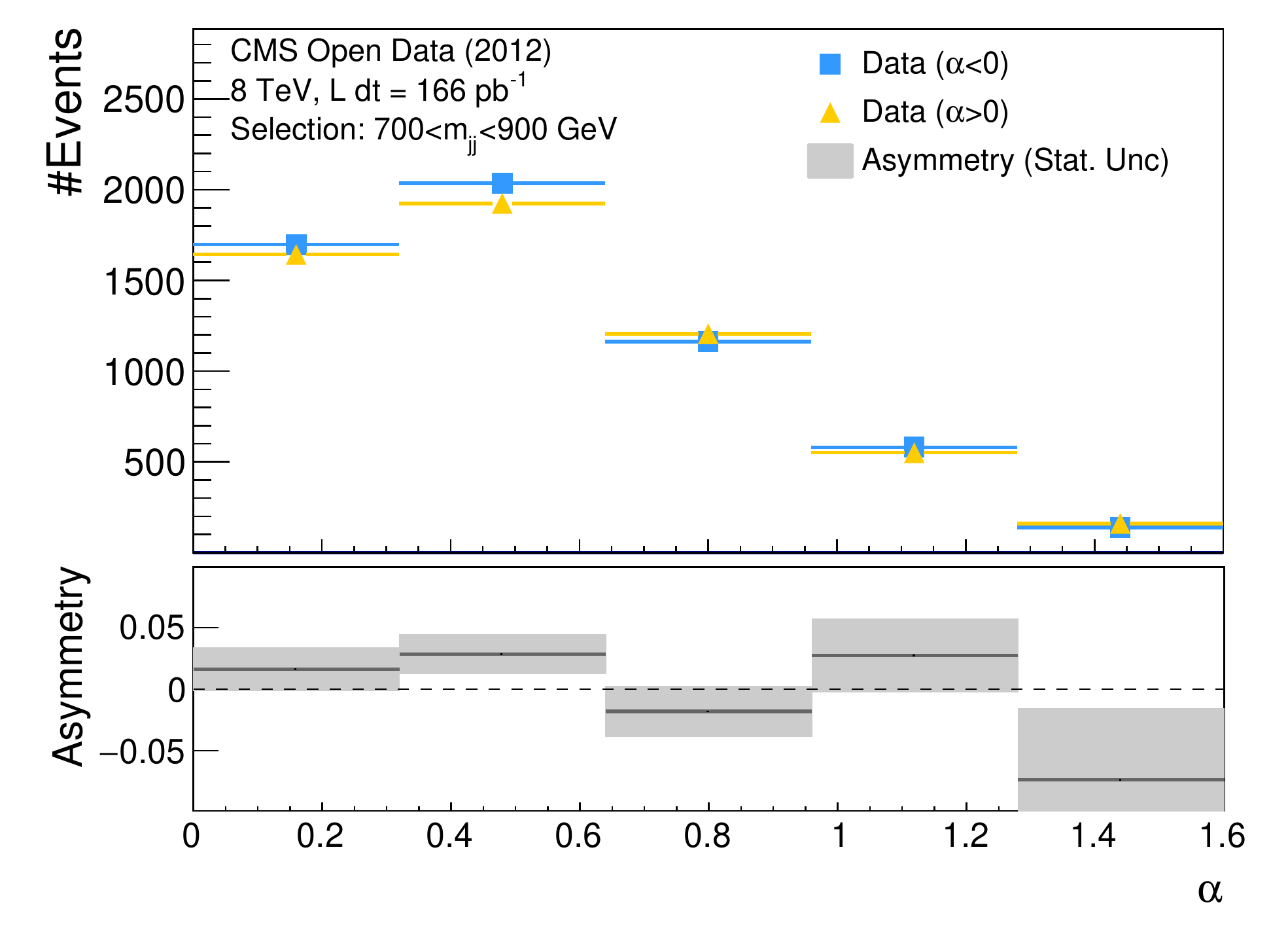}
\hspace{0.1cm}
\includegraphics[width=7.3cm]{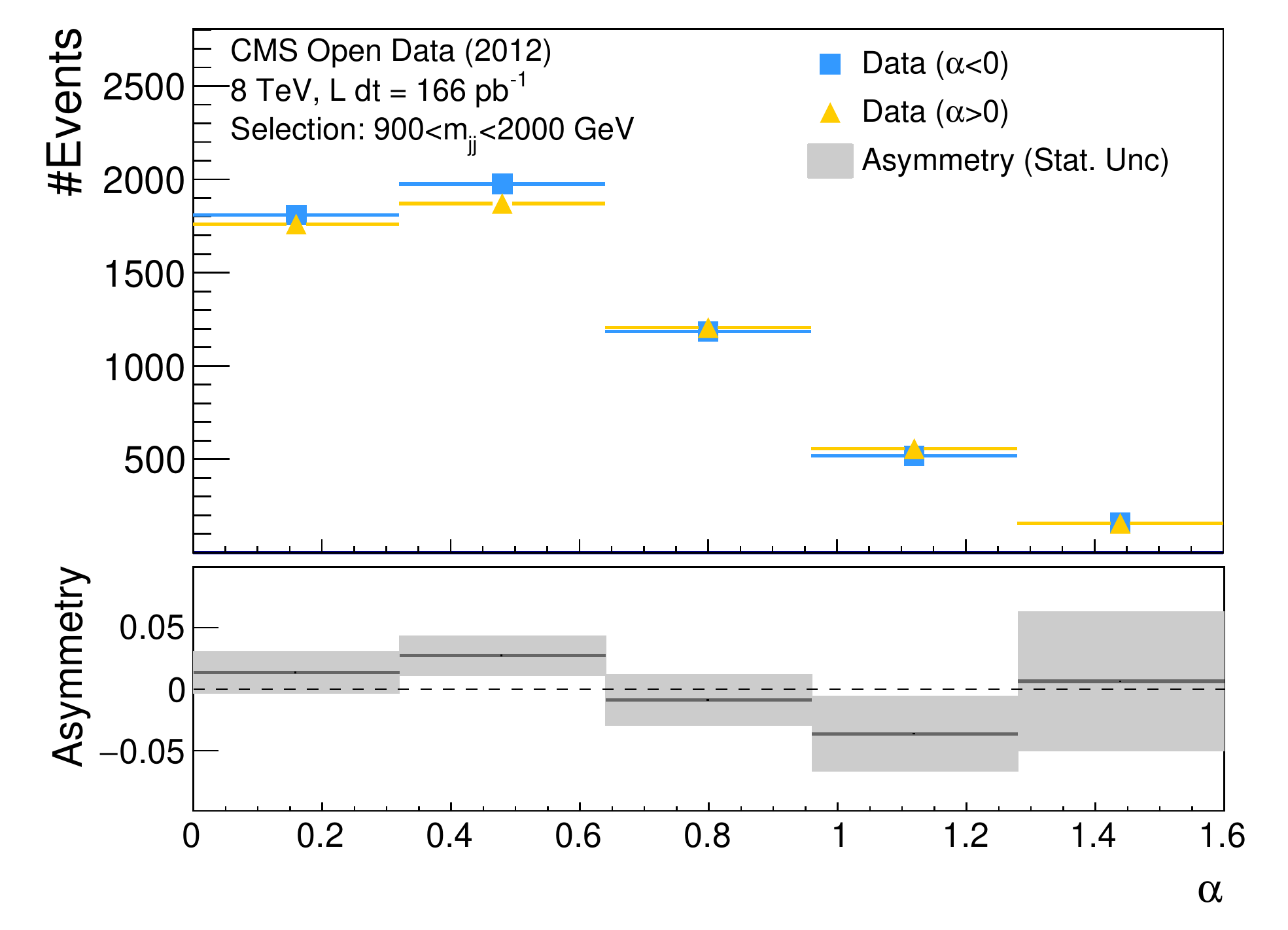}
\caption{\label{fig:AsymmetryReco} Distribution of reconstructed events with $\alpha<0$ and $\alpha>0$ for all events (upper right), as well as three regions in the invariant mass $m_{12}$; upper right: $0<m_{12}<700$ GeV, lower left: $700<m_{12}<900$ GeV, lower right: $900<m_{12}<2000$ GeV. The ratio in each plots shows the asymmetry value including statistical uncertainties.}
\end{center}
\end{figure}

\begin{figure}[tb]
\begin{center}
\includegraphics[width=7.3cm]{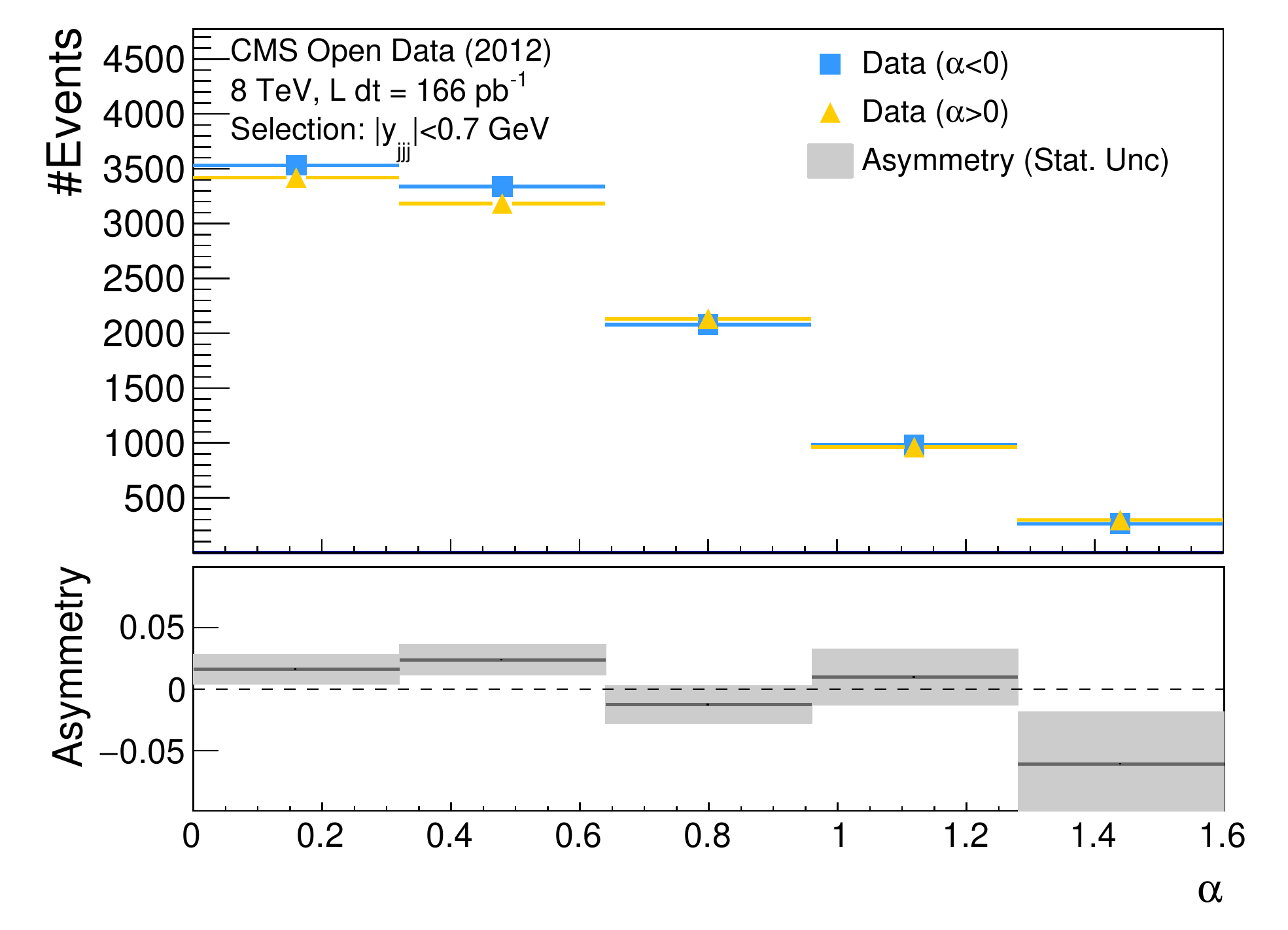} 
\hspace{0.1cm}
\includegraphics[width=7.3cm]{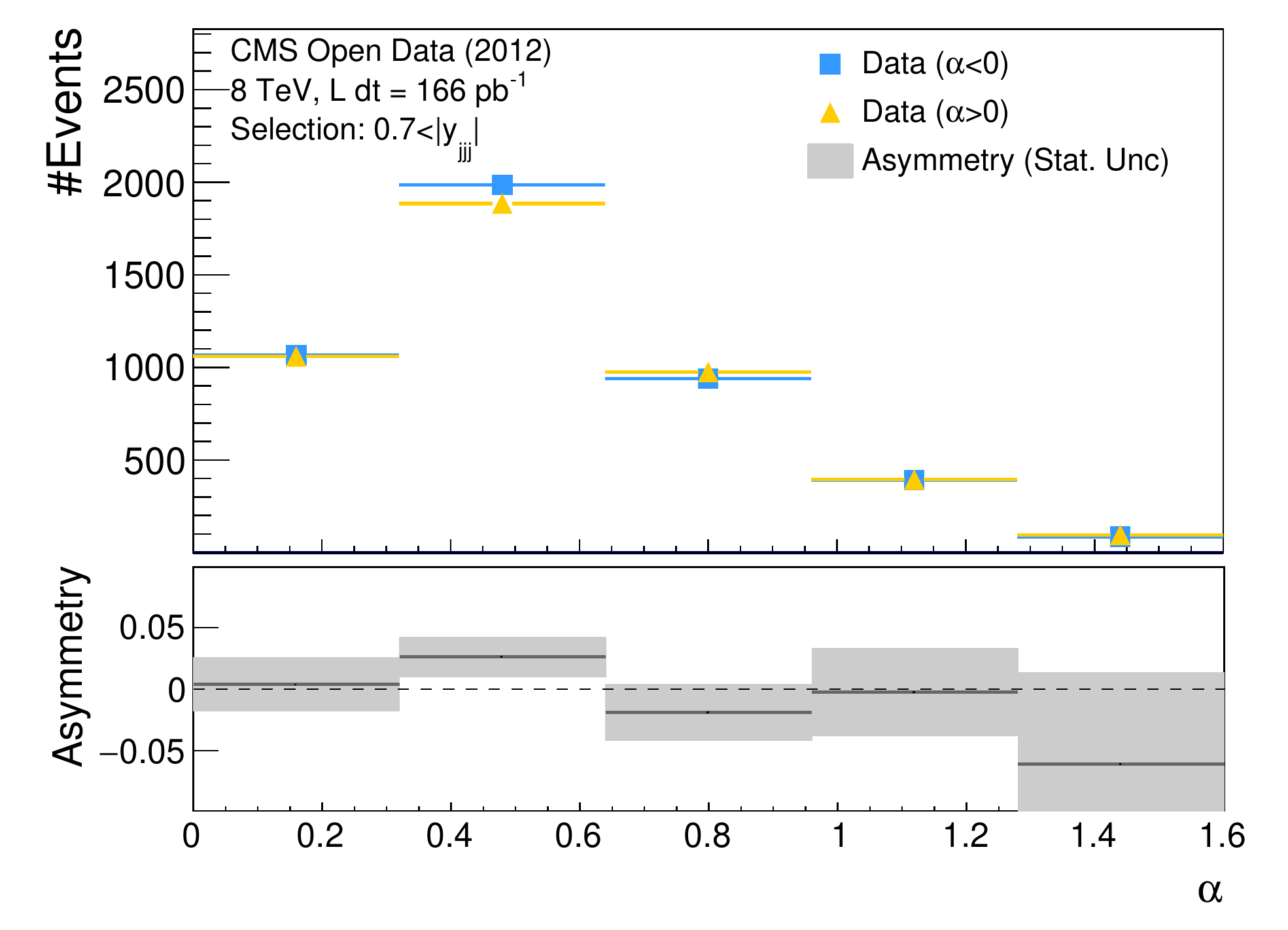}
\caption{\label{fig:AsymmetryRecoY} Distribution of reconstructed events with $\alpha<0$ and $\alpha>0$ for events with $|y_{jjj}|<0.7$ (left) and $0.7<|y_{jjj}|$ (right). The ratio in each plots shows the asymmetry value including statistical uncertainties.}
\end{center}
\end{figure}

A partial asymmetry can be calculated for different regions of $\alpha$, for example by comparing the number of events in $[ -0.96, -0.64]$ to the number of events in $[0.64, 0.96]$. This is graphically illustrated for the inclusive selection as well as the three $m_{jj}$ mass-ranges and two $y_{jjj}$ rapidity ranges in Figure \ref{fig:AsymmetryReco} and Figure \ref{fig:AsymmetryRecoY}, respectively. While the upper part of each plot shows the actual number of reconstructed events with positive and negative values of $\alpha$, the lower part shows the resulting partial asymmetry including the statistical uncertainty. In general a good agreement with the SM expectation is seen and no deviation above $2.5\sigma$ is observed.

\section{Results Corrected \label{sec:Correction} for Detector Effects}

Jets have also been  defined at the generator level, by clustering final-state particles with decay length $c\tau$ > 10 mm, using the anti-$k_t$ algorithm \cite{Cacciari:2008gp} with radius parameter R = 0.5. The generator-level jets are used to define the fiducial volume of this measurement, by selecting events fulfilling the same kinematic requirements at the generator level as at the reconstruction level. 

The normalized three-jet production cross-section, is measured as a function of $\alpha$ in the binning $[-1.60, -1.28, -0.96, -0.64, -0.32, 0.00, 0.32, 0.64, 0.96,1.28, 1.60]$.\footnote{The bin boundaries are at exactly at the decimal values listed. Since $\alpha$ cannot exceed $\pi/2\approx 1.57$ there are no overflow bins.} 

\begin{figure}[t]
\centering
\begin{minipage}{7.3cm}
  \centering
  \includegraphics[width=1.0\linewidth]{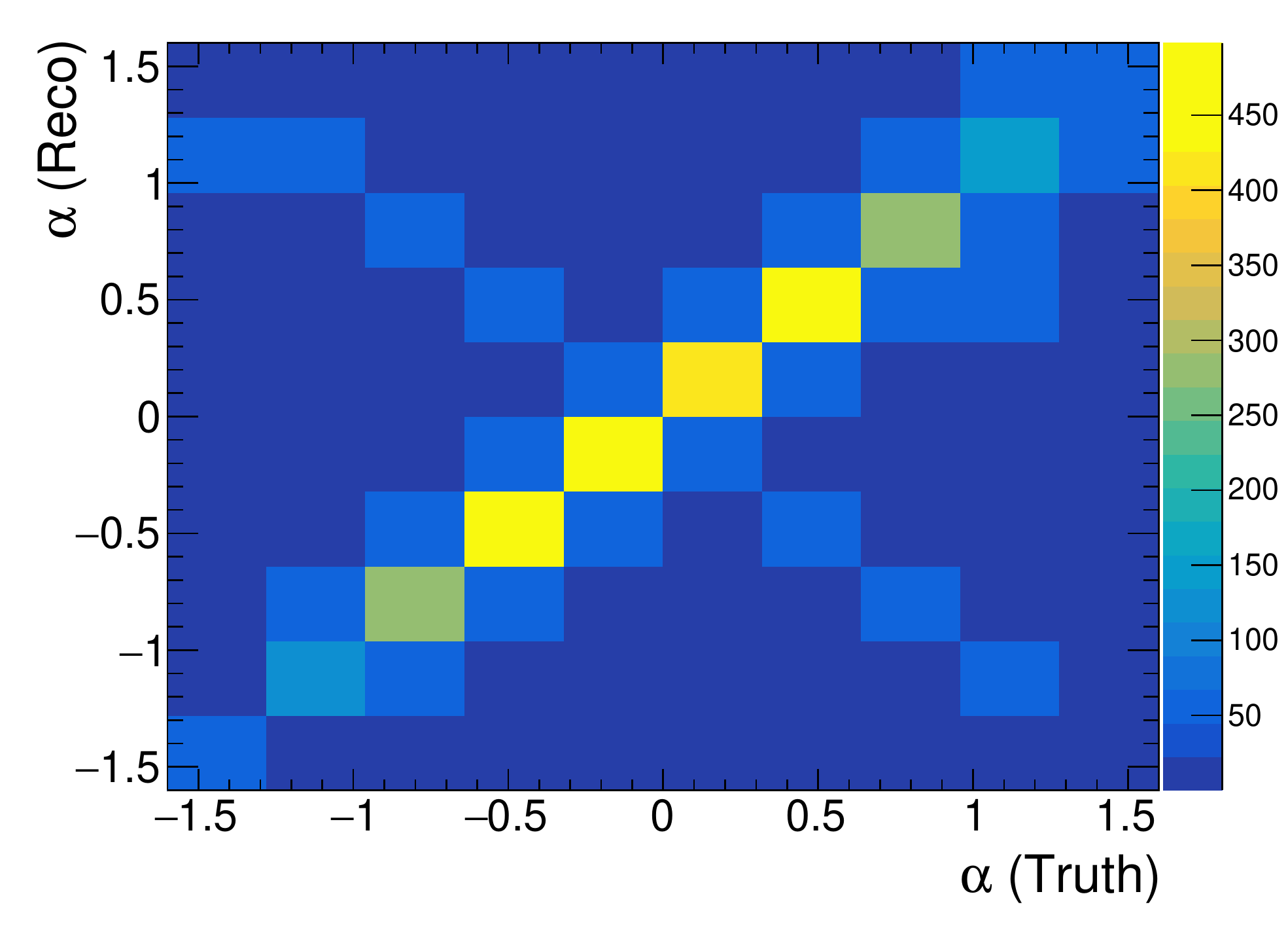}
	\caption{Unfolding Matrix used in the iterative unfolding procedure, where the truth $\alpha$ values are shown on the x-axis and the corresponding reconstructed values on the y-axis..\label{fig:UnfoldingMatrix}}
\end{minipage}%
\hspace{0.1cm}
\begin{minipage}{7.3cm}
\centering
\includegraphics[width=1.0\linewidth]{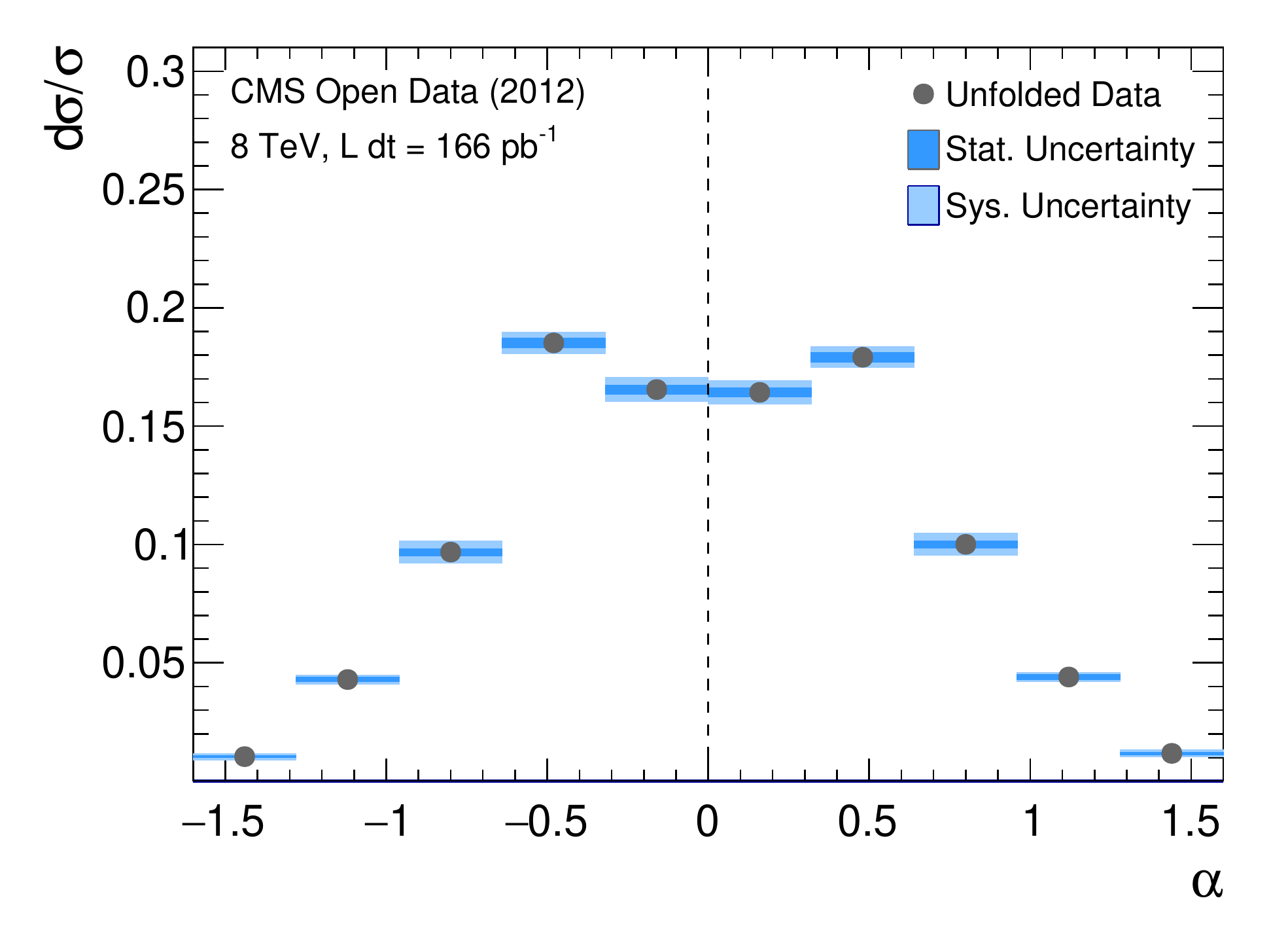}
\caption{Unfolded, detector-corrected $\alpha$ distribution with statistical and systematic uncertainties. For a conservative interpretation, the uncertainties should treated as uncorrelated.\label{fig:UnfoldedResults}}
\end{minipage}
\end{figure}

The normalized differential distribution within the fiducial volume is corrected for detector effects and bin-to-bin migrations using an iterative Bayesian unfolding method, with two iterations~\cite{DAgostini:1994fjx, DAgostini:2010hil}. First, the data are corrected for background contributions as well as events that pass the detector-level selection but not the particle-level selection. Then, the iterative Bayesian unfolding technique is used as a regularized matrix inversion to correct for the detector resolution in events that pass both the detector-level and particle-level selections. The response matrix (Figure~\ref{fig:UnfoldingMatrix}) which connects the distribution at reconstruction and particle level is estimated using the inclusive QCD MC sample. After the application of the response matrix, a final correction is applied to account for events that pass the particle-level but not detector-level selection. The chosen binning of $\alpha$ ensures a purity of over 90\% in each bin. 

Statistical uncertainties are estimated by toy variations of the input data statistics, systematic uncertainties are estimated by varying each source of uncertainty and repeating the full unfolding procedure. For the studies of the systematic effects of JES and JER related uncertainties, several scenarios are considered: in the full-correlation scenario, we apply the systematic uncertainties equally to all reconstructed jet candidates of each event. In the no-correlation scenario, we apply the systematic variations to each jet individually, but leave the other jet candidates of the event unaffected\footnote{Hence three variations of the results are derived, corresponding to the individual changes to each of the three jets}. The raw-calibration scenario is based on jet kinematics, without any applied calibrations. The difference to the nominal results is calculated for each scenario and taken as systematic. The systematic uncertainties due to the angular resolutions are estimated by adding an additional smearing on the reconstructed $\eta$ and $\phi$ values, following Section \ref{Sec:Calibration}, and then repeating the unfolding procedure. The modeling bias of the unfolding procedure has been tested in two different ways. First, the MC truth level distribution of $\alpha$ was reweighted to the data distributions and then the unfolding procedure repeated. Second, the 4-jet QCD sample was used as a MC prior. The differences between the nominal and the alternative result was taken as systematic uncertainty and added in quadrature. 

The normalized unfolded distributions of $\alpha$ are shown in Figure~\ref{fig:UnfoldedResults} together their overall uncertainties. The numerical values, including a detailed split-up of the associated uncertainties is given in Table \ref{tab:AlphaResultsUnfolded}. The total uncertainty ranges between 3\% for the central bins to 13\% for the bins corresponding to large absolute values of $\alpha$. While the central bins are dominated by uncertainties of the jet energy scale systematics, the bins for $|\alpha|>1$ are dominated by the limited statistics of data and MC as well as modeling uncertainties. It is interesting to note, that these modeling uncertainties are not due from migration effects of $\alpha$, rather than from efficiency corrections that have to be applied between the reconstruction and truth fiducial volumes.

\begin{table}[tb]
\footnotesize
\begin{center}
\begin{tabularx}{\textwidth}{l |c |c |c |c |c |c |c |c |c}
\hline
Bin		& Value		& Stat.	& MC-Stat.	& Model	& Jet 		& Jet 	 	& Jet $\eta$	& Jet $\phi$	& Total	\\
		&			& 		& 			& 		& (Uncor.)		& (Cor.)		& Res.		& Res.		&	\\
\hline
(-1.60,-1.28)	& $0.010\pm0.001$		& 4.3\%	& 4.9\%		& 7.1\%	& 7.2\%		& 5.3\%			& 2.5\%		& 0.8\%	& 13.5\%	\\
(-1.28,-0.96)	& $0.043\pm0.002$		& 2.2\%	& 2.2\%		& 0.5\%	& 2.2\%		& 2.8\%			& 0.4\%		& 1.1\%	& 5.0\%	\\
(-0.96,-0.64)	& $0.097\pm0.005$		& 1.5\%	& 1.4\%	& 3.3\%	& 2.8\%	& 0.0\%	& 0.5\%	& 0.6\%	& 4.9\%\\
(-0.64,-0.32)	& $0.185\pm0.005$		& 1.1\%	& 1.0\%	& 0.7\%	& 1.0\%	& 1.7\%	& 0.3\%	& 0.1\%	& 2.6\%\\
(-0.32,0.00)	& $0.165\pm0.005$		& 1.1\%	& 1.2\%	& 1.7\%	& 1.9\%	& 0.8\%	& 0.3\%	& 0.1\%	& 3.2\%\\
(0.00,0.32)	& $0.164\pm0.005$		& 1.1\%	& 1.1\%	& 1.7\%	& 1.9\%	& 0.8\%	& 0.3\%	& 0.1\%	& 3.2\%\\
(0.32,0.64)	& $0.179\pm0.005$		& 1.1\%	& 1.0\%	& 0.7\%	& 1.0\%	& 1.7\%	& 0.3\%	& 0.1\%	& 2.7\%\\
(0.64,0.96)	& $0.100\pm0.005$		& 1.4\%	& 1.4\%	& 3.3\%	& 2.8\%	& 0.0\%	& 0.5\%	& 0.6\%	& 4.9\%\\
(0.96,1.28)	& $0.044\pm0.002$		& 2.2\%	& 2.1\%	& 0.4\%	& 2.1\%	& 2.8\%	& 0.5\%	& 1.1\%	& 4.9\%\\
(1.28,1.60)	& $0.012\pm0.001$		& 4.1\%	& 4.6\%	& 6.8\%	& 7.1\%	& 5.2\%	& 2.3\%	& 0.8\%	& 13.0\%\\
\hline
\end{tabularx}
\caption{Unfolded normalized distribution of $\alpha$ in the fiducial volume including and a summary of statistical and systematic uncertainty.\label{tab:AlphaResultsUnfolded}}
\end{center}
\end{table}


\section{\label{Sec:Conclusion}Conclusion}

We have proposed a way of constraining non-standard sources of parity violation through measurements of the angular distribution, $\alpha$, of the radiation angle of the lowest energetic jet in three-jet events, and have trialled proposal with using CMS open data proton-proton  collisions at a center of mass energy of 8 TeV.
Potential parity violating effects beyond the Standard Model could lead to an asymmetry in the $\alpha$ distribution, which can be quantified by an asymmetry parameter, $A_\alpha$, which is expected to the zero in the SM. $A_\alpha$ was measured on data in three different kinematic regions, yielding to values which are consistent with zero. No obvious experimental challenges or limitations have been encountered, hence 
the sensitivity of this measurement could be dramatically improved with higher statistics on data and a better estimate of the corresponding experimental uncertainties. The models to which it is sensitive to could be changed by moving to other variables which probe the final state only, such as the variables $\delta$ and $S_1$ to $S_4$ defined in equations (\ref{eq:delta}) to (\ref{eq:sfour}). The limitations of the unfolded distribution can be overcome by larger MC samples as well as a more careful evaluation of the associated model uncertainties. We therefore encourage LHC collaborations to perform similar studies in upcoming full Run-2 analysis a center of mass energy of 13 TeV.


\section*{Acknowledgements}

We would like to thank the CMS collaboration for providing the full 2012 data-set as well as for the documentation on the CMS detector performance. This work would have not been possible without the excellent performance of the LHC as well as the existing computing infrastructure and the support from CERN. M.S.~would like to thank, in addition, the Fulbright commission as well as the Volkswagen Foundation for the support of this work. Moreover, he would like to thank his previous colleagues at MIT, in particular Aram Apyan, Philip Harris and his host Markus Klute for answering all questions regarding the treatment of the CMS Open Data for this project as well as the pleasant environment during the Fulbright research scholarship. C.G.L.~acknowledges fruitful and related discussions with members of the Cambridge Supersymmetry Working Group: notably with Sophie Renner, Thibaut Mueller, Ben Nachman and Thomas Gillam in 2013, and more recently with Zachary Hulcher and Rupert Tombs.


\bibliographystyle{unsrt}
\bibliography{paper}


\end{document}
\endinput